\documentclass[sigconf]{acmart}	

\usepackage{xcolor}
\usepackage{booktabs}
\usepackage{threeparttable}
\usepackage[compact]{titlesec}

\AtBeginDocument{%
  }

\usepackage{amsmath,amsfonts,algorithm}
\usepackage[noend]{algpseudocode}

 \usepackage{mdframed}
 \newmdenv[
  backgroundcolor   = gray!15 ,
  hidealllines      = true,
  innerleftmargin   = 0pt,
  innerrightmargin  = 0pt,
  innertopmargin    = 5pt,
  innerbottommargin = 10pt,
  skipabove         = .5\baselineskip,
  skipbelow         = .5\baselineskip
  ]{myframe}

\renewcommand\footnotetextcopyrightpermission[1]{}
\setcopyright{none} 
\acmConference[]{}

\begin{document}

\newcommand{\paragraphbe}[1]{\vspace{0.75ex}\noindent{\bf  #1}\hspace*{.3em}}
\newcommand{\INDSTATE}[1][1]{\STATE\hspace{#1\algorithmicindent}}
\newcommand{\gruteser}[1]{{\textcolor{red}{[gruteser: \textbf{#1}]}}}
\newcommand{\adriag}[1]{{\textcolor{red}{[adriag: \textbf{#1}]}}}
\newcommand{\kairouz}[1]{{\textcolor{blue}{[kairouz: \textbf{#1}]}}}
\newcommand{\peter}[1]{{\textcolor{blue}{[kairouz: \textbf{#1}]}}}
\newcommand{\bonawitz}[1]{{\textcolor{red}{[bonawitz: \textbf{#1}]}}}
\newcommand{\mllm}[1]{{\textcolor{red}{[mllm: \textbf{#1}]}}}
\newcommand{\ebagdasa}[1]{{\textcolor{blue}{[ebagdasa: \textbf{#1}]}}}

\title{Towards Sparse Federated Analytics: Location Heatmaps under Distributed Differential Privacy with Secure Aggregation}

\author{Eugene Bagdasaryan}
\authornote{The algorithm has been developed while the author was at Google.}
\affiliation{%
  \institution{Cornell Tech}
}
\email{eugene@cs.cornell.edu}

\author{Peter Kairouz}
\affiliation{%
  \institution{Google}
}
\email{kairouz@google.com}

\author{Stefan Mellem}
\affiliation{%
  \institution{Google}
}
\email{mllm@google.com}

\author{Adri{\`a} Gasc{\'o}n}
\affiliation{%
  \institution{Google}
}
\email{adriag@google.com}

\author{Kallista Bonawitz}
\affiliation{%
  \institution{Google}
}
\email{bonawitz@google.com}

\author{Deborah Estrin}
\affiliation{%
  \institution{Cornell Tech}
}
\email{destrin@cs.cornell.edu}

\author{Marco Gruteser}
\affiliation{%
  \institution{Google}
}
\email{gruteser@google.com}


\begin{abstract} 
{We design a scalable algorithm to privately generate location heatmaps over decentralized data from millions of user devices. It aims to ensure differential privacy before data becomes visible to a service provider while maintaining high data accuracy and minimizing resource consumption on users' devices. To achieve this, we revisit distributed differential privacy based on recent results in secure multiparty computation, and we design a scalable and adaptive distributed differential privacy approach  for location analytics. Evaluation on public location datasets shows that this approach successfully generates metropolitan-scale heatmaps from millions of user samples with a worst-case client communication overhead that is significantly smaller than existing state-of-the-art private protocols of similar accuracy.}
\end{abstract}

\maketitle

\section{Introduction}
\label{sec:intro}

Many applications, such as those that monitor the spread of infectious diseases, traffic, or mobility, benefit from users sharing location data with a service provider. To reduce potential privacy risks inherent in such applications, we ask if it is feasible to compute aggregates at scale, from on-device data of millions of participants, while ensuring differential privacy~\cite{DworkMNS06} of such aggregates before they become visible to the service provider. In particular, we seek solutions under the constraints of maintaining high data accuracy and minimizing resource consumption on user devices.

For certain statistics, differential privacy can provide reasonable accuracy in the central curator model, where a trusted party aggregates and noises the data. To eliminate the need for a trusted party, significant research has focused on the local differential privacy model (e.g., ~\cite{kasiviswanathan2008ldp, warner1965randomized, evfimievski2004privacy, duchi2013local, kairouz2016discrete}), where each participant perturbs their data to protect it even before it is aggregated at the service provider. However, the amount of perturbation needed to obtain a meaningful local privacy guarantee significantly degrades the utility compared to the central curator model.

Alternatively, secure multiparty computation (SMPC) techniques can be used to aggregate noised data in a distributed manner before the result becomes visible to the service provider~\cite{goryczka2015comprehensive}, a technique we refer to as distributed differential privacy. Since the individual contributions are cryptographically protected from other parties, each participant can add a small amount of noise that alone would not offer sufficient protection, but when aggregated with other noise shares yields the target $\epsilon$-differential privacy. Secure multiparty computation techniques, however, impose a  computational and bandwidth burden that increases with the number of participants---existing SMPC distributed differential privacy work has, to our knowledge, therefore been limited to simulations and prototypes with tens of participants~\cite{goryczka2015comprehensive}. To overcome these scaling limitations, the Honeycrisp system~\cite{roth2019honeycrisp} used homomorphic encryption that reduces the load on most users, but still requires choosing a small subset of users, the committee, to shoulder a heavy resource burden. If approximate differential privacy, rather than pure differential privacy, is acceptable, shuffling techniques also scale better in terms of the number of participants, but similarly require a trusted party or a set of outside trusted parties to conduct the shuffling task~\cite{10.1145/3132747.3132769}. None of these approaches can  collect data at scale, with accuracy comparable to differential privacy in the centralized model and without placing trust and/or a significant bandwidth or computational burden on some participants.

To address this, we revisit the distributed differential privacy concept based on recent results in the secure multiparty computation field. We develop a scalable distributed differential privacy approach applied in the context of geospatial heatmap applications at a metropolitan scale. Specifically, we build on recent breakthroughs in secure multiparty computation that can accommodate many more participants (thousands+) in vector sum computations~\cite{bonawitz2016practical,bell2020secagg}. Applying these to the distributed differential privacy concept leads to an approach that, while ostensibly straightforward, imposes several subtle challenges:

\begin{enumerate}
    \item  How should sparse data such as locations be efficiently represented so that they are compatible with secure vector sums and yield accurate, differentially private results?
    \item How should differential privacy noise be applied so that it is compatible with the integer quantization and modular arithmetic used in secure sum primitives? It's important in a practical system  when communication efficiency is a bottleneck. 
    \item How well does a secure sum primitive for thousands of participants allow geospatial statistics over millions of participants while maintaining accuracy?
    \item How resilient is the distributed differential privacy mechanism to participants that do not complete the process; for example, due to network disconnections, which inevitably arise at a larger scale?
\end{enumerate}

We tackle data representation through an adaptive histogram representation coupled with an algorithm that can determine non-uniform histogram bin sizes, or histogram meshes, even when the data distribution is unknown or changes over time. Histogram representations can be generated with low differential privacy noise (due to their low sensitivity~\cite{DworkMNS06}) and map conveniently to vector sums. The algorithm\footnote{The code is available at \url{https://github.com/google-research/federated/tree/master/analytics/location_heatmaps}.} leverages an interactive approach in which it first attempts to learn a coarse representation of the current data distribution to determine bin sizes with minimal use of the differential privacy budget. It then uses these bin sizes to query a population sample with a larger share of the privacy budget to obtain high accuracy final counts. Such an interactive approach is enabled by frameworks that can repeatedly query decentralized data from samples of user devices such as federated analytics~\cite{kairouz2019advances, RM20} and our design is grounded in experience operationalizing such solutions.

To render the results differentially private under realistic system assumptions, we first identify a distributed decomposition of the Geometric distribution. This provably yields differential privacy~\cite{goryczka2015comprehensive} under the integer quantization and modular arithmetic of the secure sum primitive, in contrast to using the more common Laplace or Gaussian noise mechanisms for differential privacy. We then devise an approach that overprovisions noise to account for clients that do not complete the protocol. It operates within a privacy budget through sets of disjoint clients and using composition over repeated queries. 

In summary, our contributions are the following:

\begin{itemize}
\item Introducing a scalable distributed differential privacy approach building
on recent secure multi-party computation advances for vector sums that yield
provable differential privacy under realistic system assumptions.
\item Proposing a novel adaptive algorithm that allows to determine communication-efficient
representations of high-dimensional sparse data (e.g., location) that are
compatible with the vector sum primitive and enhance accuracy when the data
distribution is unknown or changes dynamically. The algorithm proceeds
interactively, adapting heatmap resolution and privacy budget using results from
prior queries.
\item Evaluating the approach using simulations on public geospatial data,
demonstrating high accuracy with a worst-case communication overhead that is
significantly smaller than known private protocols of similar accuracy. 
\end{itemize}
\section{Background and Related Work}
\label{sec:appbackground}

\subsection{Applications}
\label{subsec:applications}
Private mapping of spatial distributions is a key building block for many real-world applications, thanks in part to the generality of the problem statement. A spatial distribution can be \emph{geo}spatial, i.e. representing location data on Earth. Such data are often collected by personal devices and therefore tied to the individuals that use them, leading to immediate privacy concerns regardless of the meanings of the geospatially-distributed values. Distributions of interest can also exist in other spaces, including parameter spaces of direct interest, such as a temperature-humidity ``weather space,'' or embedding spaces for arbitrary data of nearly any kind.

In this paper, we focus on public health applications in the geospatial setting, enabling epidemiologists to answer questions like:
\begin{itemize}
\item How is the population distributed over a region?
\item Where are the outbreaks of infectious disease?
\item Where are traffic patterns most deadly?
\end{itemize}without infringing unduly on user privacy.

\begin{figure}[t]
    \centering
    \includegraphics[width=1.0\linewidth]{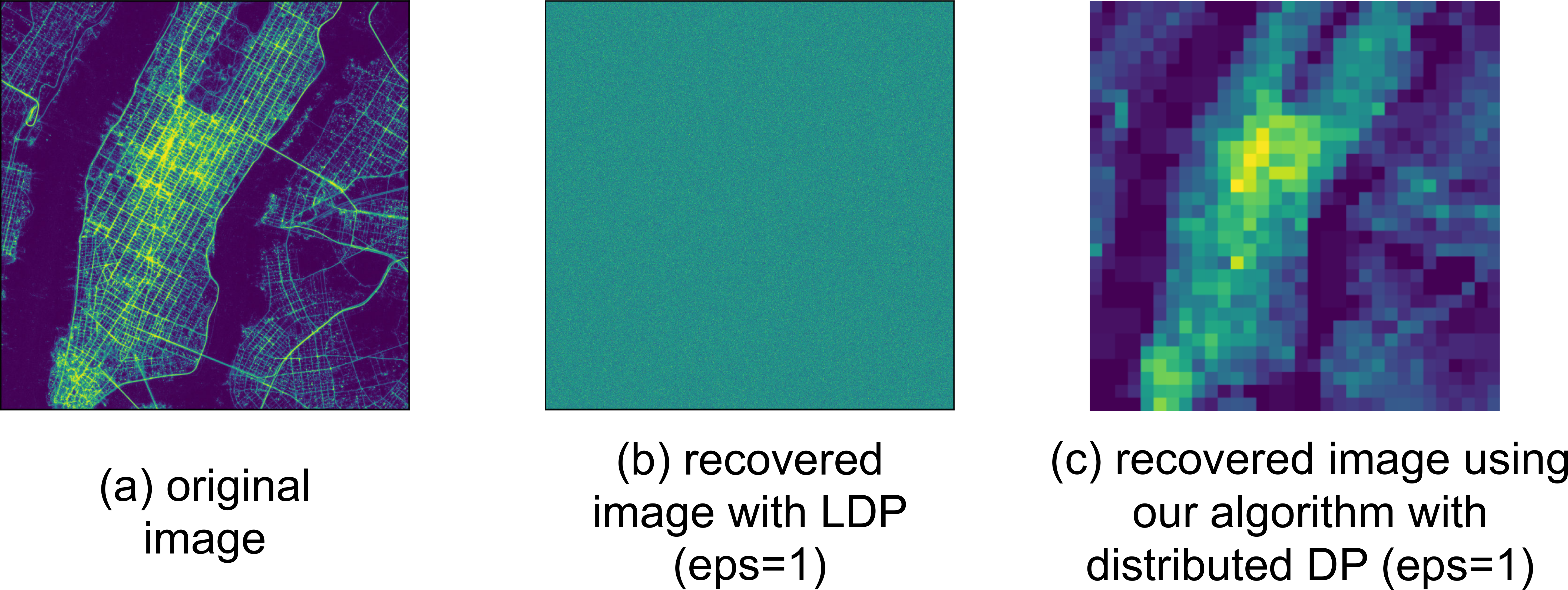}
    \caption{\textbf{Algorithms using local DP with $\epsilon{=}1$ result in high noise while our distributed DP algorithm with $\epsilon{=}1$ can recover salient heatmap characteristics}.}
    \label{fig:originalldp}
\end{figure}

Evaluating distribution measurements for epidemiological applications is a major challenge because different distributional properties are epidemiologically relevant to different diseases or policy decisions~\cite{tabataba2017framework}. For example, overestimation and underestimation are not, in general, equally problematic, nor should the relative weights of many small errors and few large errors be the same in every application. Moreover, real-world interventions must be devised with fairness and equity in mind, not just efficiency~\cite{yi2015fairness, enayati2020equity}.

Public health interventions might include emergency medical response, investment in infrastructure by governments and NGOs, distribution of limited supplies, and personal or societal behavioral alterations, and each of these can have dramatic ramifications, including significant monetary cost, disruption of livelihoods, and altered health outcomes (including lives saved or lost). In the case of a local or even individual-level intervention, such as strict social distancing, each individual decides \emph{independently} whether the risk of inaction is worth the cost of intervention. That decision can depend both on shared environmental factors, such as local disease prevalence, and on personal factors, such as the individual's medical risk factors, so there is no universal threshold at which the shared environmental factors become significant.

For simplicity, we focus our analysis on mean squared error (MSE) of the measured spatial distribution. As an idealized, motivating example, we consider a supply-distribution application in which the cost of incorrect distribution is directly proportional to MSE. To demonstrate that our algorithm is not overtuned to this metric or idealized application, we will also more briefly consider a variety of other recommended measures~\cite{tabataba2017framework} of epidemiological distribution error.

\subsection{Existing Differential Privacy Techniques}

One way to protect users' contributions is to use differential privacy (DP) ~\cite{DworkMNS06}. This method provides a rigorous, mathematical guarantee that the single user's contribution does not impact the result of a query. In the centralized model, a DP  mechanism $\mathcal{M}$ produces results from any set $\mathcal{S} \in \text{Range}(\mathcal{M}(\mathcal{D}))$ satisfying:

$$\textbf{Pr}[ \mathcal{M}(\mathcal{D}) \in  \mathcal{S} ] \leq e^\epsilon \textbf{Pr}[\mathcal{M}(\mathcal{D}') \in  \mathcal{S} ],$$
where the two databases $\mathcal{D}$ and $\mathcal{D}'$ differ by adding or removing one user's data. In the central differential privacy model, the service provider typically has access to the entire database and applies $\mathcal{M}$ to results derived from the data before release. It therefore requires trust in the service provider.

\paragraphbe{Local differential privacy} An alternative model to centralized differential privacy, which avoids the need for a trusted aggregator, is to execute the mechanism $\mathcal{M}$ at a datapoint $x\in \mathcal{D}$ before releasing it to a central service provider. Local differential privacy (LDP) ~\cite{kasiviswanathan2008ldp, warner1965randomized, evfimievski2004privacy} establishes that for all data points $x, x' \in \mathcal{D}$ and all $\mathcal{S} \in \text{Range}(\mathcal{M}(\mathcal{D}))$:

$$ \textbf{Pr}[\mathcal{M}(x) \in \mathcal{S}  ] \leq e^{\epsilon}\textbf{Pr}[\mathcal{M}(x') \in \mathcal{S} ].$$

Note that local differential privacy also implies central differential privacy, that is, it is a strictly stronger privacy model. However, a straightforward local application of the Gaussian or Laplace mechanism significantly perturbs the input and reduces utility. This is because of the compounded noise through the strict restrictions in the local differential privacy model. We demonstrate this in Figure \ref{fig:originalldp}.

\subsection{Secure Aggregation}
\label{sec:secagg}
Secure Aggregation (SecAgg) is a cryptographic secure multiparty computation (MPC) protocol that allows clients to submit masked vector inputs, such that the server can only learn the sum of the clients' vectors. In the context of federated analytics, single-server SecAgg is achieved via secret sharing (i.e. additive masking over a finite group) \cite{bell2020secagg, bonawitz2016practical,kadhe2020fastsecagg}. Clients add randomly sampled zero-sum mask vectors by working in the space of integers modulo $m$ and sampling the coordinates of the mask uniformly from $\mathbb{Z}_m$. This process guarantees that each client's masked update is indistinguishable from random values. However, when all the masked updates are summed modulo $m$ by the server, the masks cancel out and the server obtains the exact sum. 

Bonawitz et al.~\cite{bonawitz2016practical} presented the first scalable protocol that is robust to dropouts and corrupt clients possibly colluding with a dishonest server. Their protocol requires sharing pairwise masks amongst all participating clients. In practice, their protocol can support up to one thousand participating clients in one sum with vector sizes of one million~\cite{bonawitz2016practical}  since both client communication and computation costs scale linearly with the number of clients. The recent work of Bell et al.~\cite{bell2020secagg} presents a further improvement, where both client computation and communication depend logarithmically on the number of participating clients. By adopting the improved protocol, we can scale to tens or even hundreds of thousands of clients per sum. These capabilities represent several orders of magnitude improvement over earlier secure multiparty computation solutions for distributed differential privacy.
 
While the SecAgg protocol of Bell et al.~\cite{bell2020secagg} distributes
computational costs evenly across the clients,  
HoneyCrisp~\cite{roth2019honeycrisp}, a recent MPC protocol for secure aggregation, relies on a small randomly chosen 
set of clients (called committee members) for doing the heavy-lifting.
This results in a small average cost for most clients,
at the expense of steep costs for the selected committee members. The protocol execution involves a secure pre-processing phase to generate randomness which is responsible for much of the cost. As stated in \cite{roth2019honeycrisp}, this results in about $5$ minutes of compute and about $3$ GB of bandwidth consumption for the committee, seriously limiting the applicability of the protocol to mobile devices.

\subsection{Distributed DP with SecAgg}

While local DP avoids the need for a fully trusted aggregator, it leads to poor utility in practice (see Figure \ref{fig:originalldp}) given its known lower bounds such as the $O(\sqrt n)$ error for summation~\cite{kasiviswanathan2008ldp, duchi2013local, kairouz2016discrete}. To overcome this challenge without fully trusting the server, distributed DP can be used. Under distributed DP, clients first add noise to their data locally and then submit their noisy data to a private aggregation protocol. This ensures that the server only sees the (differentially private) noisy sum of the updates. 

Much of the recent work on distributed DP focuses on the shuffled model of DP where the noisy client updates are shuffled together (i.e. anonymized) before the server can see them  \cite{erlingsson2019amplification, bittau17prochlo, cheu2019distributed, GKMP20-icml, anon-power, ghazi2019private, ghazi2020pure, ishai2006cryptography, BalleBGN19, balle_merged, balcer2019separating, balcer2021connecting, girgis2020shuffled, girgis2021shuffled, wang2019improving}. With the exception of the work by Bittau et. al \cite{bittau17prochlo}, where the shuffler is instantiated securely using a trusted execution environment (TEE), these works focus on analyzing privacy-accuracy-communication trade-offs under an ideal shuffler without instantiating the exact means to implement the shuffling step. In contrast, our work proposes a complete system design including a concrete \textit{single-server secure aggregation} mechanism that builds on the secure multiparty computation protocol summarized in Section \ref{sec:secagg}. A single-server based protocol makes deployment easier compared to protocols that assume non-colluding parties (e.g. \cite{wang2019improving}) because it reduces the need for coordination with external parties. Further, any shuffling-based solution relies on generic privacy amplification results (e.g. \cite{feldman2020hiding}), which may be optimal \textit{order wise} but cannot match the exact privacy-accuracy trade-offs obtained under central DP. In contrast, under SecAgg, we can characterize the exact distribution of the noise upon summation, which in practice results in better accuracy than going through generic privacy amplification.

The combination of SecAgg and distributed DP in the context of federated analytics is far less studied. Indeed, the majority of existing works ignored the scalability, finite precision, and modular arithmetic challenges associated with SecAgg \cite{goryczka2013secure, truex2019hybrid, valovich2017computational}. This is especially constraining at low SecAgg bit-widths (e.g., in settings where communication efficiency and scalability are critical). We bypass these constraints by presenting a distributed integer-based mechanism and show how the modular arithmetic associated with SecAgg does not impact the privacy guarantees and has little effect on utility, even when clients contribute multiple locations. 

Contemporary work by Huang et al.~\cite{huang2021frequency} introduces a similar concept that combines multi-party computation with local differential privacy. However, their approach only considers batches of 100 participants and relies on sketching, which does not reach the accuracy and communication efficiency of our algorithm as we show in Section \ref{sec:experiments}. 

In the context of training machine learning models, the recent works of \cite{kairouz2021distributed, agarwal2021skellam} show how distributed discrete analogs of the Gaussian mechanism can be carefully analyzed and used to train high quality models with communication-constrained secure aggregation. These distributed discrete mechanisms achieve approximate differential privacy and work well in the context of interactive model training where privacy budgeting has to happen throughout thousands of training rounds. Our work focuses on learning location heatmaps, and we seek pure differential privacy guarantees.

\subsection{DP Histograms and Heatmaps}

Prior work on differentially private grids for geospatial data~\cite{cormode2012differentially, qardaji2013differentially}
uses adaptive mechanisms for exploration, but assumes central DP and a fixed schedule for the privacy budget across algorithm levels. Approaches such as PrivTrie~\cite{wang2018privtrie} and LDPart~\cite{zhao2019ldpart} use an adaptive threshold but a fixed privacy budget allocation whereas we investigate the adaptive privacy allocation profile.
PrivTree~\cite{zhang2016privtree} also uses adaptive thresholding in a 
setting analogous to ours, but targets the central model, and is difficult to port to the distributed differential privacy setting.
Doing so would require a secure implementation of non-linear operations such as thresholding and max. These are much less efficient than secure summation, which suffices for our solution.

The idea of using tree-like data structures for estimating histograms over large domains or discovering heavy hitters has been explored before in \cite{Cormode2003, bassily2017practical,zhu2020federated}. However, the work of \cite{Cormode2003} predates differential privacy (i.e. does not offer any DP guarantees) and the TreeHist and Bitstogram algorithms\cite{bassily2017practical} are non-interactive, rely on sketching, achieve local DP via randomized response \cite{warner1965randomized}, and assume the existence of public randomness. Our approach is interactive, does not use sketching or offer local DP, and does not require public randomness. Indeed, our work is most closely aligned with the TrieHH algorithm in \cite{zhu2020federated}, an adaptive algorithm for learning heavy hitters with differential privacy. Yet, our work is different in several non-trivial ways: (a) TrieHH relies on random sampling and thresholding to achieve approximate central DP guarantees, we explicitly use a distributed integer noise mechanism compatible with secure aggregation and focus on pure central DP; (b) TrieHH allows for only extending the leaf nodes at the lowest level of the tree, we allow the dynamic expansion and collapse of leaf nodes at all levels of the tree; (c) TrieHH uses a fixed privacy schedule across all sub-queries, we use a dynamic privacy scheduling scheme. 

More broadly, we note that there is a rich body of theoretical work on distribution learning, frequent sequence mining, and heavy-hitter discovery both in the central and local models of DP \cite{bhaskar2010discovering, bonomi2013mining, diakonikolas2015differentially, xu2016differentially, zhou2018frequent, kairouz2016discrete, wang2017locally, bassily2017practical,  acharya2018communication, ye2018optimal,  Bun2018, cormode2018marginal}. As discussed previously, the central model of DP assumes that users trust the service provider with their raw data while the local model avoids this assumption but incurs a steep loss in accuracy. Our work bridges these existing models of privacy in that it allows an honest-but-curious service provider to learn histograms and heatmaps in a centrally differentially private way without observing the users' data.

\section{Problem Definition and Threat Model}
\label{sec:definition}

Given a central server $S$ and a set of clients $\mathcal{N}$ (with $|\mathcal{N}|$ in the millions) in a metro area, each possessing a weighted set of locations $D_i\sim\mathcal{D}$ where $\|D_i\|_1=1$, the task is to estimate the overall spatial distribution $\mathcal{D}$ of users under distributed differential privacy. In particular, the goal is to maximize accuracy under a privacy budget constraint $\epsilon_{total}$ while minimizing the worst-case client communication overhead. We detail these goals, constraints, and assumptions below.

\paragraphbe{Accuracy and task variants} In particular, we focus on user density estimation (counting users in regions) and proportion estimation (measuring regional rates such as the fraction of the population satisfying some private property). These two use cases differ in two key ways: the structure of data that must be aggregated and the quantification of accuracy. In terms of data structure,  a single count per region suffices for density estimation, while proportion estimation requires two counts per region (hinting toward generalizations with $k$ counts). In terms of accuracy, we seek to optimize distance-from-ground-truth measures such as mean squared error (MSE) or $L_p$ norms for density estimation and mean confidence interval size for proportion estimates. Appendix~\ref{subsec:more_metrics} further expands on different metrics that are useful for specific applications.

\paragraphbe{Privacy threat model} We consider an adversary with, {\em simultaneously}, observation access to the aggregation server and full control over a small fraction of compromised clients. That is, the adversary can trace the execution and observe the state at the server, and tamper with the execution of compromised clients. The objective of the adversary is to infer the exact user location of clients that the adversary does not control. Out of scope are active adversaries that can compromise and gain access to the set of clients outside their control or adversaries that seek to poison the results.

\paragraphbe{Assumptions} For simplicity, we start by assuming that each user reports only a single location $D_i=d_i\in\mathcal{D}$ (e.g., their home address). Our approach scales similarly when each user supplies a set of locations. We detail that extension in Section \ref{subsec:multi_loc} and include results in Table \ref{tab:multi_loc}.

We assume that users who report their location will reveal their IP address during the data collection protocol. In many cases, this can imply location at a coarse level such as country or even city, but not the fine-grained locations we seek to protect. We further assume that a significant number of users (e.g. 100K or more) occupy areas that are smaller and more private than can be resolved via IP geolocation.

We assume that clients can be programmed to transform the data before transmission to the server (e.g., encode location coordinates into a vector or generate noise) and have enough resources to run these transformations and the secure aggregation protocol (see Section~\ref{sec:experiments} for overhead). We further assume that the server supports interactive communication with clients and that the server (and clients) can run the secure aggregation protocol with the size of each aggregation being limited to $S_{max} << |\mathcal{N}|$ clients ($S_{max}$ being on the order of thousands whereas $\mathcal{N}$ is on the order of millions).

We also assume that the secure aggregation protocol implementation is correct and the server cannot tamper with the program code. However, some portion of the clients $q<S_{max}/2$ in a single shard can drop out or be malicious. We assume that the remaining majority of clients are correct and argue that it would be difficult to create a botnet of sufficient size in one metro area to break this assumption. 

\section{Distributed Differentially Private Histograms}
\label{sec:systemdesign}
Before we present our end-to-end algorithm for learning complex heatmaps, let us develop a basic distributed differential privacy algorithm for learning histograms over closed, known domains and show how it can handle the challenges discussed in the introduction: (a) the integer modulo arithmetic associated with secure aggregation, (b) secure aggregation shards with a maximum of thousands (as opposed to millions) of clients, and (c) client dropouts (i.e. clients that are admitted to a shard but never report back). Even though the solution we present in this section can be used as is to learn complex, largescale heatmaps, we will show, in the next section, that this approach will not provide good utility and will come at steep communication and computation costs. To overcome these challenges, we extend this design through an adaptive hierarchical histogram algorithm that efficiently encodes user location to increase accuracy and reduce communication overhead. 

The most natural way to learn a histogram over locations with a secure aggregation protocol is to have participating clients ``one-hot'' encode their location into a vector of length equal to the total number of possible locations\footnote{This assumes that the set of all possible locations is finite, as could be achieved by quantizing the exact locations into a fixed grid of locations. We expand on this in Section~\ref{subsec:baselines}.}. With such a representation, participating clients can simply submit their one-hot encoded vectors to a secure aggregation protocol that ensures that the service provider only gets to see the sum, which is essentially a histogram over all possible locations.

\begin{figure}[t!]
    \centering
    \includegraphics[width=0.8\linewidth]{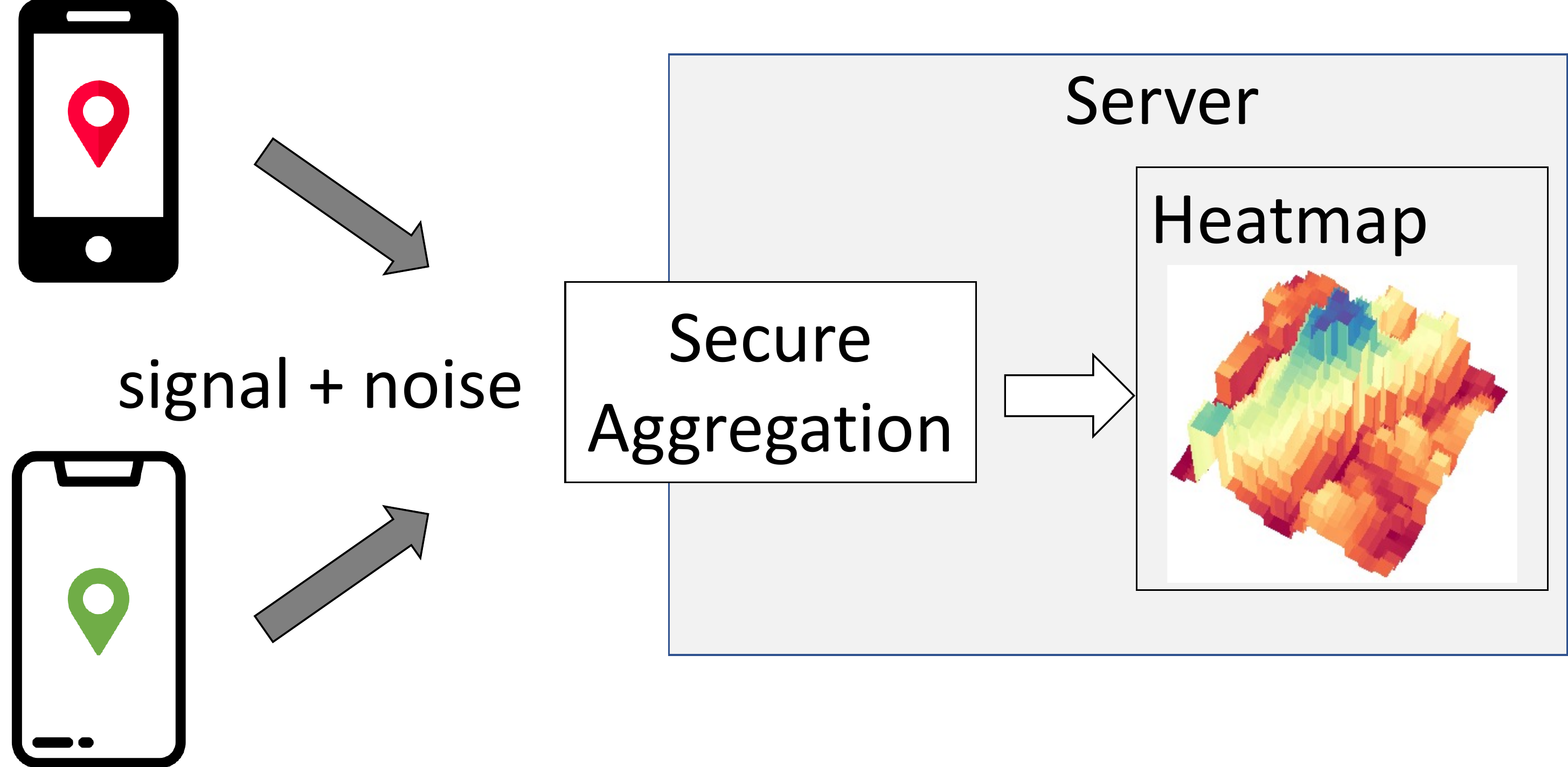}
    \caption{\textbf{Overview of reporting location data with distributed differential privacy using secure aggregation}.}
    \label{fig:overview}
\end{figure}

\paragraphbe{Achieving distributed differential privacy under integer modulo arithmetic} 
While the above approach ensures that the server cannot learn the locations of participating clients and only observes a histogram over all possible locations, it does not offer any rigorous privacy guarantee. To achieve such a guarantee, participating clients can add noise locally in a way that makes the noisy sum differentially private (see Figure~\ref{fig:overview}). However, because secure aggregation operates over integers modulo $m$ arithmetic this rules out techniques that add continuous (floating point) noise, including the widely used Gaussian or Laplace mechanisms. We bypass this issue by having clients: (a) add integer noise drawn from the difference of two independent P\'{o}lya random variables to their one-hot encoded vectors, and (b) modulo clip the entries of the noisy vector (i.e. apply entry-wise modulo $m$) before submitting the clipped noisy values to a secure aggregation protocol.

Our approach is summarized in Algorithm \ref{alg:1}. For the available set of users $\mathcal{N}$ we pick a subset $U$ and compute a histogram as a vector that we later project to a map using some encoding $vmap$ (we use a quadtree that maps 1-dimensional vector to a 2-dimensional map).
We now analyze the end-to-end differential privacy guarantees. 

\begin{definition}[P\'{o}lya Random Variables]
A random variable  $X$ is a P\'{o}lya random variable with parameters $(\alpha, \beta)$ if it has a probability mass function given by
\begin{equation}
P (X = k) = {\alpha + k - 1 \choose \alpha - 1} \beta^k (1 - \beta)^\alpha,
\end{equation}
for $k \in \mathbb{N}$ and ${\alpha + k - 1 \choose \alpha - 1}$ defined using Gamma functions. We denote this distribution by P\'{o}lya$(\alpha, \beta)$. To draw $X \sim$ P\'{o}lya$(\alpha, \beta)$, we first draw $\lambda \sim Gamma(\alpha, \beta/(1-\beta))$, where $Gamma(.)$ is the Gamma distribution, and then use $\lambda$ to draw $X \sim Poi(\lambda)$, where $Poi(.)$ is the Poisson distribution.
\end{definition}

\begin{theorem}[Distributed Discrete Laplace Mechanism]
Assume $X_{i}$ and $Y_{i}$ are independent P\'{o}lya$(\alpha,\beta)$ random variables for $i \in \{1, \cdots, S_{max}\}$. Let  $\alpha = 1/S_{max}$ and $\beta = e^{-\varepsilon/\Delta}$. Then the following securely aggregated query
\begin{equation}
\label{eq:mod_clip}
\left(\sum_{i=1}^{S_{max}} \left(vector_i + X_{i} - Y_{i}\right) \mod m \right) \mod m,
\end{equation}
where $vector_i$ represents client $i$'s encoded integer vector with $L_1$ norm bounded by $\Delta$, is $\varepsilon$ differentially private. 
\end{theorem}

The proof of the above theorem hinges on the following two observations. First, 

\begin{equation}
Z = \sum_{i=1}^{S_{max}} X_{i} - Y_{i}
\end{equation}

is distributed according to 

\begin{equation}
    \label{eq:geopmf}
    P(Z = k) = \frac{1 - e^{-\varepsilon/\Delta}}{ 1 + e^{-\varepsilon/\Delta}} e^{-\varepsilon|k|/\Delta},
\end{equation}

which is a discrete version of the Laplace noise that achieves $\varepsilon$ differential privacy when added to queries with $L_1$ sensitivity equal to $\Delta$ \cite{ghosh2012universally}. See Theorem 5.1 in \cite{goryczka2015comprehensive} for a detailed proof of this claim for $\Delta =1$. Second, the modulo clipping operation can be viewed as post-processing to an already differential private query. To view this, observe that $(\sum x_i \mod m) \mod m = (\sum x_i) \mod m$. This means that the modulo clipping step applied by each client commutes with the modulo sum, thus appearing as a post-processing to a query that was noised with discrete Laplace noise. This proves that our approach achieves pure differential privacy irrespective of the choice of secure aggregation precision $m$. Having said that, $m$ does play a critical role in determining the accuracy of our approach -- aggressive modulo clipping saves bandwidth and computations but introduces more bias. We investigate the values of $m$ that preserve accuracy in Section \ref{subsec:secagg_effect}.

\begin{figure}[t]
    \vspace*{-\baselineskip}
    \begin{minipage}{\columnwidth}
     \begin{algorithm}[H]
        \caption{Basic Histogram estimation with secure aggregation}
        \label{alg:1}
        \begin{algorithmic}[1]
        \State \textbf{Parameters:} available users $\mathcal{N}$, number of users queried per histogram $U$, privacy budget $\epsilon_{total}$, SecAgg shard size $S_{max}$, SecAgg precision $m$, location encoding map $vmap$.
        \State \textbf{Constants:} expected dropout rate $\delta_{drop}{=}0.95$, query sensitivity $\Delta{=}1$

        \vspace{0.2cm}
        
                \Function{Histogram}{$\epsilon, \mathcal{N}, U, S_{max}, vmap$}
                \State \textit{\# derive noise parameters using Eq.~\ref{eq:stdgeo}}
        \State $\alpha=1/((1-\delta_{drop}) \cdot S_{max})$
        \State $\beta = e^{-\varepsilon/\Delta}$
        \State $ shards = U / S_{max}$
         \For{SecAgg shard $r$ \textbf{in} $shards$}
            \State $u_r \leftarrow $ sample $S_{max}$ from $\mathcal{N}$ (no replacement)
            
            \State \textit{\# Run secure aggregation on clients $u_r$}
             \State $hist_r  = $ \textbf{SecAgg}(method=\textbf{ClientUpdate}, args=$(\alpha, \beta, vmap)$, users=$u_r$)
          \EndFor
         \State $hist = \sum_{r=1}^{shards} hist_r$
        \State map = \textbf{project\_to\_map}$(hist, $vmap$)$
        \State \texttt{return} $hist, map$
        \EndFunction
        
        \vspace{0.2cm}
        \Function{ClientUpdate}{$\alpha, \beta$, $vmap$}
        \State $location = \textbf{retrieve\_location}()$
        \State $vector = \textbf{encode}(location, $vmap$)$
        \State $vector = \textbf{ModuloClip}(vector + \textbf{Noise}(\alpha, \beta), m)$
        \State \texttt{return} $vector$
        \EndFunction
        
        \vspace{0.2cm}
        \Function{Noise}{$\alpha$, $\beta$}
        \State $X, Y \sim$ P\'{o}lya$(\alpha, \beta)$
        \State \texttt{return} $X - Y$
        \EndFunction
        
        \vspace{0.2cm}
        \Function{ModuloClip}{$noisy\_vector$, $m$}
        \State \texttt{return} $noisy\_vector \mod m$
        \EndFunction
        
        \end{algorithmic}
     \end{algorithm}
    \end{minipage}
\end{figure}

\paragraphbe{Multiple aggregation shards} In order to reduce communication costs and make client cohort assembly easier, secure aggregation shards are typically limited in size, say to $S_{max} \leq 10,000$ clients~\cite{bonawitz2016practical}. The above routine can be run multiple times separately (shown on line 6 of Algorithm \ref{alg:1}) to expand the reach of the algorithm. And privacy budgeting is not required in this case because different shards are run on disjoint sets of users. Nevertheless, the noise added to each shard compounds across shards. Shard size thus plays a crucial role, mediating a tradeoff between accuracy and communication costs (as well as cohort assembly difficulty). We investigate the effect of the shard size on the accuracy of the learned heatmaps in Section \ref{subsec:secagg_effect}.

\paragraphbe{Handling dropouts and adversarial clients }
The above approach gives provable pure differential privacy guarantees as long as all participating clients follow the protocol faithfully until the end. In practice, a small fraction of clients may be adversarial or drop out in the middle of a secure aggregation shard. If unaddressed, the server observes an inaccurate sum of noise shares (due to dropped out or malfunctioning clients). Therefore, the partial sum of noise shares is not statistically equivalent to sampling from a discrete Laplace distribution, thus reducing the privacy guarantees. To account for this issue, we propose scaling up the standard deviation of the P\'{o}lya random variables used to generate the local noise shares. For example, if we anticipate a maximum of 5\% noise dropout rate $\delta_{drop}=0.05$, then in the worst case, we end up with $0.95S_{max}$ participating clients (as opposed to $S_{max}$ clients) and call it \textit{DP shard size}. This means that the $\alpha$ parameter of the  P\'{o}lya random variables should be scaled as  $\frac{1}{(1-\delta_{drop})S_{max}}=1/(0.95S_{max})$ as opposed to $1/S_{max}$. 
Finally, secure aggregation fails to give the correct sum when a substantial fraction of the clients, more than $\delta_{drop}$, drop out \cite{bonawitz2016practical, balle2018improving}. This means that we can design for this worst-case scenario to ensure that either the server sees a provably differentially private sum or nothing. 
See Appendix~\ref{subsec:secagg_effect} for an analysis of the impact of noise dropouts on the resulting quality of learned heatmaps. Adversaries that seek to poison the heatmap, for example by adding excessive noise, are out of scope for this work.

\begin{figure*}[tbp!]
    \centering
    \includegraphics[width=0.95\linewidth]{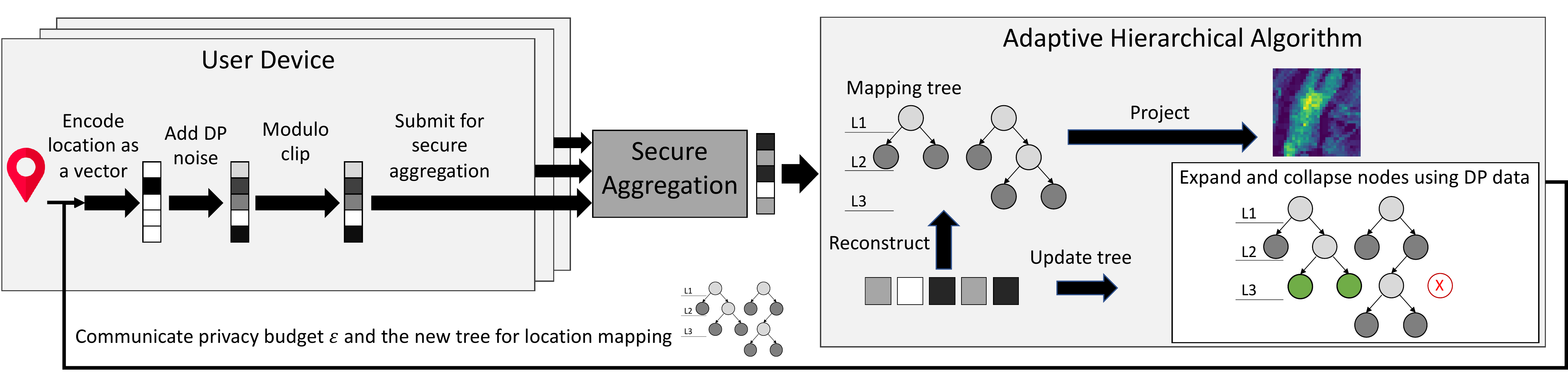}
    \caption{\textbf{System design}.}
    \label{fig:main}
\end{figure*}

\section{Adaptive Hierarchical Histograms}
\label{sec:algorithm}

The basic primitive presented in the previous section can be used as-is to learn complex heatmaps. However, as we show in Table \ref{tab:expresults} of Section \ref{sec:experiments}, this simple approach fails to provide high accuracy when the underlying heatmap is large. To overcome this issue, we now extend the basic primitive through an adaptive hierarchical histogram algorithm that executes a variable-resolution differentially private histogram query through a series of sub-queries, whose parameters (both domain and privacy budget allocation) are determined by the result of the previous sub-queries. Each sub-query can contain many shards, in which clients' data is queried via secure aggregation.

\subsection{Baselines and Their Shortcomings}
\label{subsec:baselines}
Let us first consider known approaches to further understand the trade-offs between accuracy and communication overhead. For simplicity, let us assume that each client only reports one location.

\paragraphbe{Flat one-hot encoding} In this approach, we apply Algorithm~\ref{alg:1}'s basic histogram estimation to a single spatial histogram over regional cells of uniform spatial dimension. Given a target area (say, a city metro area) and a desired maximum spatial resolution, we define a uniform raster of $D$ cells and associated $vmap$ (the mapping from locations to vectors in Algorithm \ref{alg:1}). Clients determine which raster cell their location falls into, encode it into a one-hot vector of size $D$, add differential privacy noise, and send this vector to the server via secure aggregation as described in Algorithm~\ref{alg:1}.

This approach thus incurs communication costs proportional to $D$, and its accuracy is very sensitive to the data distribution: if the raster is too fine, resulting in a small number of users in each cell, then their signal is overpowered by differential privacy noise. On the other hand, making the raster coarse reduces $D$ and thus communication overhead, but does not allow accurate estimation of the distribution over space.

\paragraphbe{Hierarchical one-hot encoding} Before running the algorithm, we define a resolution hierarchy, such as the $L$ resolution levels $P \times P$, $2P \times 2P$, $...$, $2^{L-1}P \times 2^{L-1}P$. Note that each regional cell at level $\ell$ is composed exactly of $4$ cells at level $\ell+1$. Each client divides their privacy budget $\varepsilon_{total}$ evenly among the $L$ resolution levels and runs the flat one-hot encoding algorithm at each level, optionally appending the $L$ one-hot encodings together before sending the results to the server via a single secure aggregation. The server then selects, for each subregion of the target area, the resolution whose result maximizes map accuracy, discarding data from the unused resolutions.

This approach resolves the problem of overpowering a high resolution signal with noise by allowing the server to select a resolution with a good signal-to-noise ratio during postprocessing. However, it spends privacy budget inefficiently by collecting and then discarding data from the other $L-1$ resolutions. Moreover, it does not reduce the communication overhead noted above; in fact, the cost is even larger (by a constant factor that depends on the arity of the resolution hierarchy).

\paragraphbe{Count--min (CM) sketch hierarchical encoding}
The previous two approaches require communication costs that scale with the size $D$ of the target area's raster(s). This is a problem in settings where $D$ is much larger than the number of clients $S_{max}$ who are involved in each aggregation. CM sketches address precisely this problem by using $O(S_{max}\log S_{max})$ space instead of $O(D)$ space as in the two previous algorithms. This algorithm proceeds exactly like the previous one except that it encodes each of the intermediate $L$ histograms as a sketch, and thus trades $O(D)$ communication for $O(L S_{max}\log S_{max})$ communication, which is beneficial for large $D$.

Nevertheless, since clients still transmit location information associated with every resolution and the server only selects one resolution per subregion, discarding the rest of the data, this approach retains the core wastefulness of the hierarchical one-hot encoding algorithm, failing to improve its accuracy.

\subsection{Leveraging Interactivity with Adaptive Hierarchical Histograms}

Our algorithm employs a series of sub-queries to interactively partition and re-partition the spatial map into cells of different sizes, each with a good signal-to-noise ratio (see Figure~\ref{fig:main}). We leverage this interactivity by introducing adaptivity with regard to: (i) adapting the spatial resolution in each sub-query and (ii) adapting the privacy budget allocated to each sub-query. 

\paragraphbe{Adaptive resolution} By utilizing interactivity, our algorithm can avoid collecting data from the too-fine regions that would be discarded during postprocessing by the non-interactive approaches outlined above, thus reducing the communication costs drastically without impacting accuracy. We represent the spatial partition as a quaternary tree in which each node represents a regional cell and each child node represents a quadrant of its parent's region. At each iteration, clients report their data to the lowest node in the tree whose corresponding region contains their location, i.e. by determining the longest matching prefix in the tree structure. The server aggregates the reported locations and then considers the signal to noise ratio of each node to decide whether to split that region into its four children, collapse it into its parents, or retain it for the next iteration.

Naively, the server could split and collapse based on constant, data-independent thresholds for the number of required reports in a subregion. We can improve upon this by considering the factors that limit the optimal resolution: sampling noise and differential privacy noise. In Section \ref{subsec:adares}, we introduce a heuristic adaptive mechanism to maximize accuracy by selectively splitting or collapsing regions based on the data from prior sub-queries and both sources of expected noise.

\paragraphbe{Adaptive privacy} In an interactive protocol, each sub-query incurs a privacy cost, raising the question of how the overall privacy budget $\varepsilon_{total}$ should be allocated over the sequence of interactive sub-queries. When the algorithm starts without a good prior for the actual distribution, the initial sub-queries are too coarse or too fine to capture the actual client distribution and can be viewed as queries to optimize the final query partition rather than measure client density within that partition.

In Section \ref{subsec:adapriv}, we introduce an adaptive privacy budget schedule to spend only as much privacy budget as is necessary on the initial resolution-optimization queries and to allocate as much privacy budget as possible to the final density-estimation query using an appropriate partition.

\subsection{Adapting Spatial Resolution}
\label{subsec:adares}

Given a starting partition, the algorithm executes a sub-query and uses its results to adapt the resolution for the next sub-query. In particular, it splits cells if the differentially private count is sufficiently high that a finer resolution is feasible. If the count is too low on a cell, it also has the option to revert back to coarser cells. It then generates a new query and repeats the process. Note that the algorithm makes these decisions independently on each cell, which can result in a non-uniform partition with finer resolution in some areas and coarser resolution in others. The algorithm terminates when the finest possible resolution is reached in each cell as determined by an expansion criterion and then reports final counts for this partition.

The algorithm stops dividing a cell into finer cells when there is a low probability to learn more about the spatial distribution within the cell, given the expected differential privacy noise and finer-cell counts. More intuitively, the algorithm stops when further cell splitting would render the counts small enough that they become indistinguishable from noise. 

Let $\varepsilon_{total}$ be the total privacy budget for the map and let $\varepsilon_\ell$ be the privacy loss incurred during subquery $\ell$. The remaining budget at sub-query $j$ is then $\varepsilon_{rem} =\varepsilon_{total} - \sum_{\ell=1}^{j-1} \varepsilon_\ell$. Assuming the discrete Laplace differential privacy mechanism, we can use this remaining budget to estimate a lower bound for the noise standard deviation at any remaining sub-query as follows: 

\begin{equation}
    \label{eq:stdgeo}
    \sigma_{rem}= \tau(\varepsilon_{rem}, \Delta)= \sqrt{2\beta/(1-\beta)^2},  
\end{equation}
where $\beta=e^{-\varepsilon_{rem}/\Delta}$.

We can learn more by splitting cell when the signal from users (e.g., count) significantly exceeds the expected noise. Given $n_i$ users observed in a cell $i$ during the previous sub-query $j-1$, we expect $n_i$ users spread over the four sub-cells if we split this cell during the next sub-query and the maximum expected count is $n_i$ (if they were all located in the same sub-cell). We therefore heuristically set the stopping criterion as

\begin{equation}
    n_i > k \sigma_{rem}
\end{equation}

where k is a parameter than can be adjusted. We use $k = 2$ standard deviations in our implementation, meaning that the best case expected $n_i$ should exceed the 95th-percentile of the noise lower bound.  

A similar criterion can also be used to collapse cells when the signal becomes indistinguishable from noise to restore higher accuracy. In the context of our partition tree $\mathcal{T}$, we collapse cells by simply deleting the corresponding node. This means that if some leaf nodes contain counts indistinguishable from noise, they can be recombined in the following sub-queries. 

We allow nodes to be partially split, i.e. only a subset of the four child nodes are present. Users report their data at the lowest node (longest matching prefix) in the tree that covers their raw location, so if a node has just one child with sufficient signal, the data from the remaining three child regions get aggregated together. If these remaining regions associated with the parent node still have insufficient signal, that node can be collapsed again so that its signal will be aggregated into \emph{its} parent region. It is therefore important to choose a representation of $\mathcal{T}$ which permits orphaned nodes.

\subsection{Adapting Privacy Budget Schedule}
\label{subsec:adapriv}

Our algorithm aims for an accurate and fine-grained grid given a privacy budget over the entire sequence of sub-queries. One approach is simply to use a fixed schedule of the allocated budget per query based on a maximum number of sub-queries. However, the server can observe the differentially private results between the sub-queries and adjust the budget for the next sub-query based on the current state of the grid and participating users to make accurate decisions on splitting or collapsing nodes in the tree. 

 We aim to use just enough of the privacy budget for the next sub-query to make informed decisions on where to split nodes. Specifically, the expected user count per cell should significantly exceed the expected differential privacy noise. The expected user count per cell is simply $U/T$, where $U$ is the number of users that are part of the sub-query and $T$ is the number of cells and size of the securely aggregated vector (in terms of the tree $\mathcal{T}$, this is the number of nodes with fewer than four children). We accomplish this by estimating a target standard deviation of the differential privacy noise as 

\begin{equation}
\tilde{\sigma} = \tau_{target}(c, U, T, S_{max}) = c\frac{U}{T}\sqrt{\lceil S_{max}/U \rceil}
\label{eq:sigma}
\end{equation}

where $c$ is a calibration parameter (typically $c<1$) and $S_{max}$ is the maximum number of users in one secure aggregation shard. Calibration parameter $c$ can depend on the shape of  distribution (uniform or heavily concentrated in certain areas). Lower values of $c$ results in accurate splits but fewer queries (good for uniform distributions) and higher values of $c$ results in more queries but lower split accuracy (good for concentrated cases). Furthermore, setting the correct value of $c$ depends on the task, e.g. discovering the most accurate location of few hotspots might need to increase $c$, whether generally accurate map might need lower $c$ value. In practice, the optimal $c$ can be estimated on similar inputs (e.g. maps of other cities) or prior runs of the algorithm.

The last term in Equation~\ref{eq:sigma} accounts for noise variance accumulation when the number of users exceeds the limit on one secure aggregation shard and summing over multiple secure aggregations is necessary. By solving Equation~\ref{eq:stdgeo} for $\varepsilon$ we can use derived standard deviation value to spend privacy budget:

\begin{equation}
\tilde{\varepsilon} = \phi(\sigma) =-\Delta\cdot\ln(\frac{\sigma^2+1-\sqrt{2\sigma^2+1})}{\sigma^2})
\label{eq:phi}
\end{equation}

To manage the remaining privacy budget we create the following rule:

\begin{equation}
\varepsilon = \gamma(\varepsilon_{rem}, \tilde{\varepsilon}) =
\begin{cases}
    \tilde{\varepsilon}, & \text{if}\ b\tilde{\varepsilon} <= \varepsilon_{rem} \\
    \varepsilon_{rem}    & \text{otherwise}
\end{cases}
\label{eq:rem}
\end{equation}

Intuitively, we use $\tilde{\varepsilon}$ if enough privacy budget remains for a subsequent sub-query. Otherwise, we set the $\varepsilon$ to use the remainder of the privacy budget to maximize accuracy and the algorithm terminates after this last query. The expansion parameter $b$ controls how much more budget we need for the next sub-query, e.g. lower $b$ is useful when we expand only few leafs and larger $b$ when we expand all the leafs.  We then generate noise using derived $\varepsilon$ through Eq.~\ref{eq:stdgeo} and apply it to the user vector. This approach allocates more noise when the number of cells is small, such as in the initial sub-queries starting from the root of the tree. Conversely, it allocates less noise as the resolution increases.

\subsection{Algorithm and Privacy Guarantees}

Algorithm~\ref{alg:dynamic} shows the aggregate of the methods proposed above: adapting resolution and privacy budget.

\paragraphbe{Communication overhead} This algorithm significantly reduces the amount of data clients need to send to the server by optimizing the vector size. For each algorithm sub-query, a sampled client sends a vector of length equal to a number of nodes in the tree that do not have a full set of children.

\paragraphbe{Privacy Guarantees} We begin with a total privacy budget $\varepsilon_{tot}$. The algorithm combines privacy budgets across multiple sub-queries $q$ using basic composition, i.e. we spend budget $\varepsilon_q$ in the $q$-th sub-query and ensure that $\sum_q \varepsilon_q \leq \varepsilon_{tot}$. For each SecAgg shard we rely on privacy guarantees of distributed DP from Equations~\ref{eq:mod_clip},~\ref{eq:geopmf} that allow each user to add a difference between two P\'{o}lya variables to obtain discrete Laplace noise.
 
 \begin{figure}[t]
    \vspace*{-\baselineskip}
    \begin{minipage}{\columnwidth}
     \begin{algorithm}[H]
        \caption{Extending HISTOGRAM (Algorithm \ref{alg:1})}
        \label{alg:dynamic}
        \begin{algorithmic}[1]
        \State \textbf{Inputs:} available users $\mathcal{N}$, number of users queried per histogram $U$, privacy budget $\varepsilon_{tot}$,
        SecAgg shard size $S_{max}$,
         calibration $c$.

         \vspace{0.2cm}
         
        \Function{AdaptiveHierarchicalHist}{$\mathcal{N}, U, \varepsilon_{tot}$}
        \State sub-query $q{=}0$
        \State remaining budget $\varepsilon_{rem} {=} \varepsilon_{tot}$
        \State $\mathcal{T}_q = \textbf{init\_prefix\_tree()}$
        \While{$\varepsilon_{rem} > 0$ }
            \State $q = q{+}1$ \textit{\# next sub-query}
            \State \textit{\# Get target threshold using Eq.~\ref{eq:sigma}}
          \State $\tilde{\sigma_q} = \tau_{target}(c, \mathcal{T}_q, U, \varepsilon_{rem}, S_{max})$
          \State \textit{\# Get corresponding budget using Eq.~\ref{eq:phi}}
          \State $\varepsilon_q = \phi(\tilde{\sigma_q})$
          \State \textit{\# Utilize Eq.~\ref{eq:rem} to prevent overflow of $\varepsilon_{rem}$}
          \State $\varepsilon_q = \gamma(\varepsilon_{rem}, \varepsilon_q)$
          \State $\varepsilon_{rem} = \varepsilon_{rem} - \varepsilon_q$
          \State \textit{\# Use Algorithm~\ref{alg:1}}
          \State $(hist, map) = \textbf{Histogram}(\varepsilon_q,\mathcal{N}, U, S_{max}, \mathcal{T})$
          \State $\mathcal{T}_{q+1} = \textbf{UpdateTree}(\mathcal{T}_q, hist, \tilde{\sigma_q})$
        \EndWhile

        \State \texttt{return} $(hist, map)$
        \EndFunction
        
        \vspace{0.2cm}

         \Function{UpdateTree}{tree $\mathcal{T}$, histogram $hist$, threshold $t$}
         \ForAll{$node \in \mathcal{T}$}
            \State $count = hist[node]$
            \If{count>$t$}
            \State $node.split()$
            \EndIf
            \If{count<=$\frac{1}{4} t$}
            \State $node.collapse()$
            \EndIf
         \EndFor
         \State \textbf{return} $\mathcal{T}$
         \EndFunction
        
        \end{algorithmic}
     \end{algorithm}
    \end{minipage}
\end{figure}

 Note that, while at each sub-query $q$ we can run multiple randomly constructed shards of secure aggregation,
 these are all disjoint, and therefore regardless of a number of shards the algorithm only spends $\varepsilon_q$ budget. This fact, combined with the security of the underlying Secure Aggregation protocol $\Pi$, implies that an attacker that is consistent with the threat model of $\Pi$ will only observe a differentially private result throughout the protocol execution. This end-to-end privacy guarantee of Algorithm 2 is stated informally in the following lemma, assuming that the underlying aggregation protocol is the one by Bell et al.~\cite{bell2020secagg}.

\begin{lemma}
Let $\mathcal{A}$ be an adversary controlling a  minority of the clients, 
and observing the execution trace of the server.
The view of $\mathcal{A}$ in an
execution of Algorithm~\ref{alg:dynamic} is computationally $\varepsilon_{tot}$-DP, even if an arbitrary number of honest clients drop out.
\end{lemma}

 SecAgg relies on standard cryptographic primitives (e.g. authenticated encryption) that are defined in terms of a computationally bounded adversary. Our distributed DP guarantee depends on the security of the SecAgg protocol, and thus holds against probabilistic polynomial time adversaries. This notion is described in the literature as “computational differential privacy”~\cite{mironov2009computational}.

\begin{figure}[tbp!]
    \centering
    \includegraphics[width=0.95\linewidth]{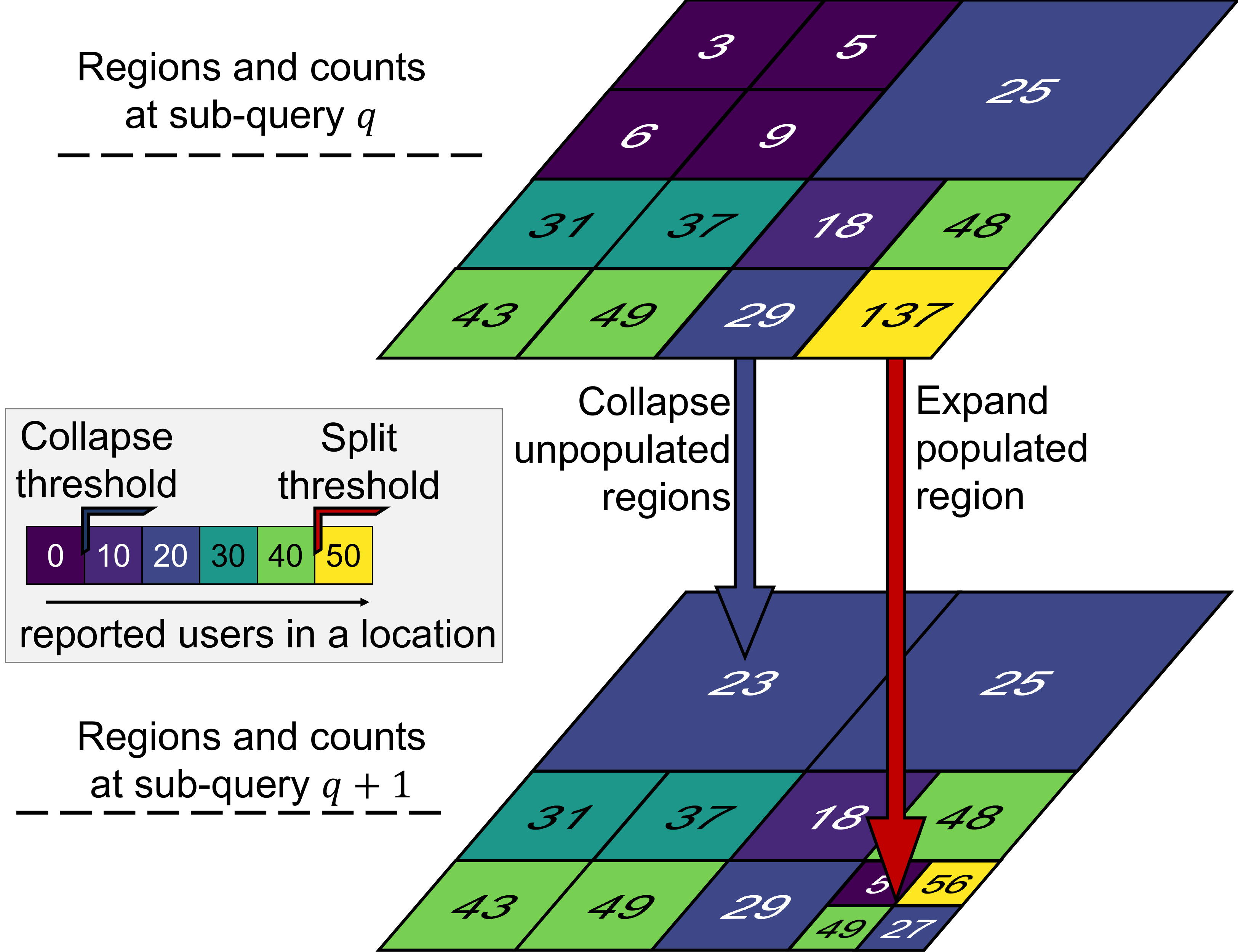}
    \caption{\textbf{Node expansion and collapse effect}.}
    \label{fig:visual_algo}
    \vspace{0.2cm}
\end{figure}
\subsection{Implementation Details}

As described above, we define a hierarchy over a space using a quaternary tree in which each node represents a rectangular region and has up to four children representing quadrants of the rectangle. Unlike the well-known quadtree~\cite{samet1984quadtree} structure, we do not require each node to be either a leaf node or have exactly four children; a node with one to three children represents all of the remaining quadrants not represented explicitly by descendants. Moreover, orphaned child nodes are permitted such that, for example, a node could have a single grandchild but no children, in which case the node represents the entire region excluding the sixteenth associated with the grandchild. All nodes with fewer than four children can accumulate counts from clients, so each such node is mapped to the index of a vector element that accumulates counts for its associated region. Other branching factors are possible and could improve performance on particular datasets. Probabilistic structures like Mondrian trees could also be used, but they require multiple iterations over data.

\begin{figure*}[ht!]
    \centering
    \includegraphics[width=1.0\textwidth]{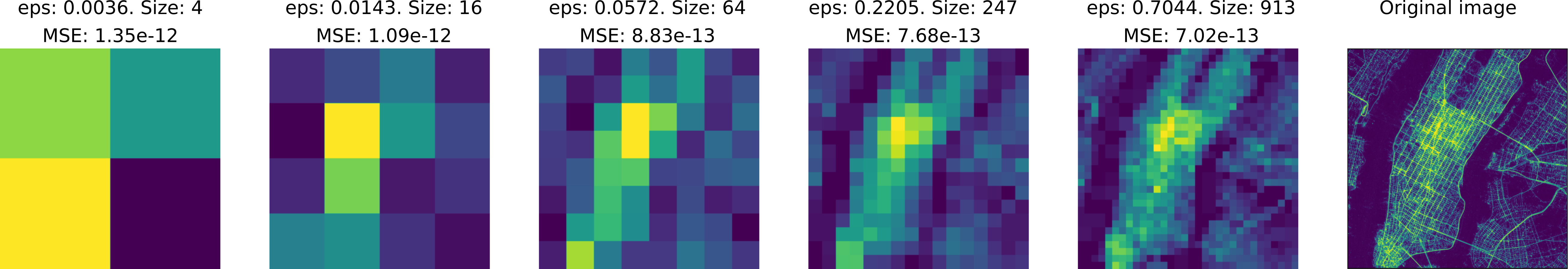}
    \caption{\textbf{Algorithm results for the Manhattan map.} We run our algorithm with $\varepsilon=1$ budget across $100,000$ users, dropout $10\%$, $S_{max}=10,000$ and size of the distributed $DP$ shard $9,000$.}
    \label{fig:results}
\end{figure*}

\paragraphbe{Transforming coordinates to a bitstring} Given a maximum depth of the tree, each location coordinate can be mapped to the deepest node of the tree that it falls into. Node IDs can be assigned in the form of a bitstring based on the node's position in the tree. The empty string represents root and, for a branching factor of 4, we traverse the tree appending at each level a two-bit suffix $\{00\}$, $\{01\}$, $\{10\}$, or $\{11\}$ corresponding to the choice of the NW, NE, SE, or SW quadrant of the parent. The resolution is thus determined by the node's depth in the tree (or, equivalently, its ID length).

For example, location $(x=12,y=5)$ on the $16\times16$ map becomes $(x=1100, y=0101)$. By grouping together the first bits for each dimension---in this case $1$ for coordinate $x$ and $0$ for coordinate $y$---we construct the first non-root ancestor node id $\{10\}$. We then append the next most significant bits from each dimension in turn such that the full sequence of node ids from largest to smallest region containing the location is $\{\} \rightarrow \{10\} \rightarrow \{10/11\} \rightarrow \{10/11/00\} \rightarrow \{10/11/00/01\}$ (with slashes included here between each quadrant suffix for readability only). When reporting location, each client contributes data to the vector element corresponding to the longest node id in $\mathcal{T}$ which matches the prefix of that client's location.


\subsection{Multiple Locations Per User}
\label{subsec:multi_loc}

Let us now consider the case where users can contribute multiple locations to a histogram. One approach is to simply adjust the sensitivity to account for multiple locations. This can be sufficient if all users contribute the same number of locations, though in practice it is common that some users visited many more locations than others.  One way to handle the latter is to limit user contributions to a maximum per user (say, one location per user) and adjust the sensitivity to this maximum. However this can add bias and significantly undermine the utility of the resulting map.

We therefore propose to preserve user-level privacy of location traces and limit each user's contribution through weighted normalization. This recognizes that many applications benefit from weighting locations according to some metric such as fraction of time spent in that location. For a given level and state of the histogram tree the user reports all weighted locations and normalizes the vector to have $L_1$-sensitivity of $1$. This results in floating point values, however, that are not compatible with secure aggregation, which operates over integers in a finite group. 

To comply with discrete methods we add a scaling constant $\gamma$ that will scale $L_1$-sensitivity and correspondingly scale the discrete geometric noise. After scaling by $\gamma$ we randomly round the resulting numbers to integers, e.g. a value $0.8$ is rounded to $0$ with probability $20\%$ and $1$ with probability $80\%$. 
For a $d$-dimensional vector that is randomly rounded there is some chance that all elements of the vector will be rounded up, changing $L_1$-sensitivity from $\gamma$ to $\gamma + d$. Therefore, we set sensitivity while computing privacy budget to $\gamma + d$.

Once the vectors from reporting users summed up in the shard we scale the result back. Note, that scaling all the values and performing secure aggregation may result in overflow during quantization (see Appendix~\ref{subsec:secagg_effect}), and requires controlling scaling $\gamma$, secure aggregation shard size, and modulo clipping.

 This method achieves pure DP. In particular, the devices: (1) form an all zero vector representing all possible locations they could ``vote on'' depending on the tree learned so far, (2) increment the counts of locations they are in, (3) normalize the vector by its $L_1$ norm (i.e. divide by the total number of locations they voted on), (4) scale the vector by a large number $\gamma$, (5) stochastically round each entry of the vector to the nearest integer. If the vector has length $d$, then the normalized, scaled, and rounded vector will \emph{always} have an $L_1$ norm$\leq \gamma +d$. Therefore, we can use the distributed discrete Laplace mechanism with a sensitivity of $\gamma +d$ to ensure pure DP. Also, upon securely summing the devices' noisy vectors, the server can divide back by $\gamma$. This ensures that the extra blowup in noise variance (due to the rounding step) is at most a factor of $(1 + d/\gamma)$. 
 Therefore, we choose $\gamma$ to be much larger than $d$ to mitigate the blowup, i.e. $1+d/\gamma \approx 1$.
\section{Evaluation}
\label{sec:experiments}

We evaluate our design on public location data in terms of density estimation accuracy and communication cost.

\subsection{Experimental Setup}

The algorithms are implemented using NumPy and Tensorflow Federated~\cite{TFF2019} and executed on a workstation. 

\paragraphbe{Datasets} We derive location datasets from public fine-grained population heatmap images in three areas with different spatial and population distribution characteristics. In particular, we use a heatmap of Manhattan released by the New York Times~\cite{nytimes} (also used in~\cite{erlingsson2020encode}) and a heatmap of 
Mayotte and Lagos, Nigeria obtained from the Humanitarian Data Exchange~\cite{humdata} website. All heatmaps are cropped to the resolution of $1024\times1024$.

These heatmaps do not provide information per user. Therefore to derive a distribution of user locations, we interpret the luminosity value of a pixel located at the
$(x, y)$ coordinate in the image as users being located at that coordinate. Each simulated user in the generated dataset is assigned a single location. This results in $54,599,988$ users for Manhattan, $242,528$ for Mayotte, and $9,578,296$ for Lagos.

\paragraphbe{Task.} We formulate the task of reconstructing the location heatmap as a density estimation problem. We set the privacy budget to $\varepsilon = 1$.

\paragraphbe{Accuracy metrics} We focus on the mean squared error (MSE) and $L_1$ metrics---common metrics for density estimation tasks. To compute MSE, let functions $f$ (for the original map) and $f^*$ (for the algorithm's output)
return normalized density values for any coordinate $(x,y)$ in the original map. Since the resolution of the algorithm output can be coarser than the original map, the corresponding tree may not have a leaf node matching only this coordinate. In this case, we uniformly distribute the value from the deepest node that still encompasses this coordinate over the coordinates not covered by any of its children. That is, a node representing a $P \times P$ region with value $v$ and $\psi$ coordinates covered by its children would yield a value of $v/(k^2-\psi)$ for coordinates not covered by children. Given these functions, MSE is:

$$MSE(f^*, f) = \frac{1}{n^2} \sum_{x=1}^{n} \sum_{y=1}^{n} (f^*(x,y) - f(x,y))^2$$

Similarly we compute an $L_1$ distance metric. Our algorithm works on a sample of the simulated user data, and cannot fully recover the original image, even without any differentially private noise due to sampling error. Additionally, binning 
size (or tree depth) impacts the error -- some small bins might have no data while larger bins provide inaccurate count.

\begin{table*}[]
\caption{Density estimation accuracy (MSE) and communication overhead (vector size) for different user sample sizes with privacy budget $\varepsilon=1$ and secure aggregation shard size $S_{max}=10,000$. Communication overhead sums vector sizes for every algorithm level.}
\label{tab:expresults}
\centering
\begin{tabular}{lrrrrrr}
           & \multicolumn{2}{c}{$10k$ users} & \multicolumn{2}{c}{$100k$ users} & \multicolumn{2}{c}{$100k$ users} \\
Algorithm & \multicolumn{2}{c}{(no dropout)} &  \multicolumn{2}{c}{(no dropout)} & \multicolumn{2}{c}{(10\% drop)} \\
\cmidrule(r){2-3} \cmidrule(r){4-5} \cmidrule(r){6-7} & MSE (e-13) & Comm & MSE (e-13) & Comm & MSE (e-13) & Comm \\
\midrule
\textit{Non-interactive} & & \\
Non-private & $7.75$ & $2$ & $6.19$ & $2$ & $6.55$ & $2$  \\
Flat & $41.40$ &  $1,048,576$  & $21.40$ &  $1,048,576$ &  $21.93$ & $1,048,576$ \\
Hierarchical          & $8.81$           & $1,398,100$  & $7.75$ & $1,398,100$ & $7.79$ & $1,398,100$           \\
CM-sketch based & $11.62$  & $40,000$ & $11.69$ & $40,000$ & $11.73$ & $40,000$  \\
Adaptive-grid~\cite{qardaji2013differentially} & $9.67$ & $460$ & $9.89$ & $3,448$ & $9.96$  & $2,882$ \\
Spatial decomposition~\cite{cormode2012differentially} & $8.56$ & $1,048,576$ & $7.72$ & $1,048,576$ & $7.78$  & $1,048,576$ \\[2ex]
Ours (scheduled $\varepsilon$, adaptive $T$) & $7.88$ & $340$ & $6.99$ & 1254 & $7.02$ & 1244 \\               
\bottomrule
\end{tabular}
\end{table*}

\subsection{Accuracy and Communication Efficiency}

We compare our algorithm to the baselines described in Section \ref{subsec:baselines} on a number of benchmarks. A good algorithm should reconstruct the original heatmap both with low error and low communication costs while satisfying privacy constraints. The NYC map has a maximum resolution of $2^{10} \times 2^{10} = 2^{2*10}$ that corresponds to a $10$ level quadtree that can be built over the map. In our setting we allow $10,000$ clients to participate in the algorithm and aim to preserve total privacy budget of $\varepsilon=1$ per user. Users can send a vector of integers to the server using secure aggregation but can obtain the location tree in the clear as the intermediate trees are differentially private. Table~\ref{tab:expresults} shows the results.

\paragraphbe{Non-interactive baselines} We begin with a non-interactive setting where the server can only receive one response from each of the $10,000$ users.

The total number of users presented on the original map is $55$ million and sampling only $10,000$ user location will result in some inaccuracy of the result even without privacy constraints. We estimate this sampling error via a \textit{non-private baseline} in which users report their locations in plain to the server. As the server has access to all $10,000$ locations, we build a map for every level of the quadtree and pick the level that produces the best metric. The result we obtain is $MSE=7.75e^{{-}13}$ with only 2 integer coordinates communicated to the server.

With a \textit{flat one-hot encoding}, the target resolution is chosen a priori. Users report their locations at the finest resolution in our dataset via a vector of length $2^{2*10}$. This allows the addition of distributed discrete Laplace differential privacy noise to each coordinate during aggregation, but those coordinates will only accumulate signal from a few users due to their sparsity at the finest resolution, resulting in an error $41.4e^{{-}13}$ that has no meaningful result. Furthermore, this approach incurs overhead cost of $2^{2*10}$ integers per user. We cannot easily postprocess this result into a coarser resolution by summing the values at neighboring coordinates since the noise would still dominate. Alternatively, we could guess a coarser target resolution to obtain more signal at less communication overhead but risk unnecessary loss of resolution if users are clustered in a few locations. 

Under \textit{hierarchical one-hot encoding}, users report their location for every quadtree level using budget $\varepsilon=0.1$ per each level. The server analyses reported vectors and picks the best split. For this evaluation, we use the split that achieves the lowest MSE compared to the ground truth. In practice, the error would be greater due to probabilistic decisions based on expected noise. The communication overhead is a sum of coordinate vectors from $2^{2*1}$ to $2^{2*10}$, total of $1,398,100$ vector elements per user. The algorithm achieves $8.81e^{{-}13}$ MSE, only adding $15\%$ error over the non-private solution.

\textit{Count-Min sketching} can reduce the reported vector size if the client locations are sparse compared to the dimension of the aggregated quadtree levels. We define a count-min sketch with $2,000$ width and $20$ depth that achieves error below $10$ with probability $0.001$ for $10,000$ user reports. To achieve the same differential privacy guarantees we increase the sensitivity of each report proportional to the sketch's depth. Users report 10 times on different prefixes of their location using the provided sketch achieving higher error of $8.81e^{{-}13}$ but reducing communication costs to $400,000$. 

The \textit{Adaptive grid} method~\cite{qardaji2013differentially} can be considered a partially interactive algorithm since it performs two queries. The first query obtains counts on a uniform $10 \times 10$ grid. The second query is with a refined partition, where each of the aforementioned grid cells is further divided into a $n \times n$ sub-grid. Here, $n = \sqrt{\frac{N' (1-\alpha)\epsilon}{c_2}}$,  where $N'$ is a current count in that region, $\alpha$ is a budget share and $c_2$ is a hyperparameter. Following the guidelines we set $\alpha{=}0.5$ and $c_2{=}10$. To make this compatible with our distributed setting, we use our P\'{o}lya mechanism instead of the central differential privacy mechanism. This method, reduces communication by eliminating hierarchical structure, but provides lower accuracy due to its single attempt to partition the map and does not improve with more users participating in the algorithm.

We adapt differentially private \textit{spatial decomposition}~\cite{cormode2012differentially} to our distributed setting using the proposed geometric budget. This method is non-interactive and assigns more privacy budget to higher levels of the tree to obtain correct initial splits unlike our approach that saves more budget for later layers. The result of the best layer is better than the hierarchical one-hot encoding, however the algorithm does not use the threshold to split the tree and due to specifics of the budget allocation deeper layers are less accurate and consume communication budget.

\begin{figure}[t!]
    \centering
    \includegraphics[width=0.9\linewidth]{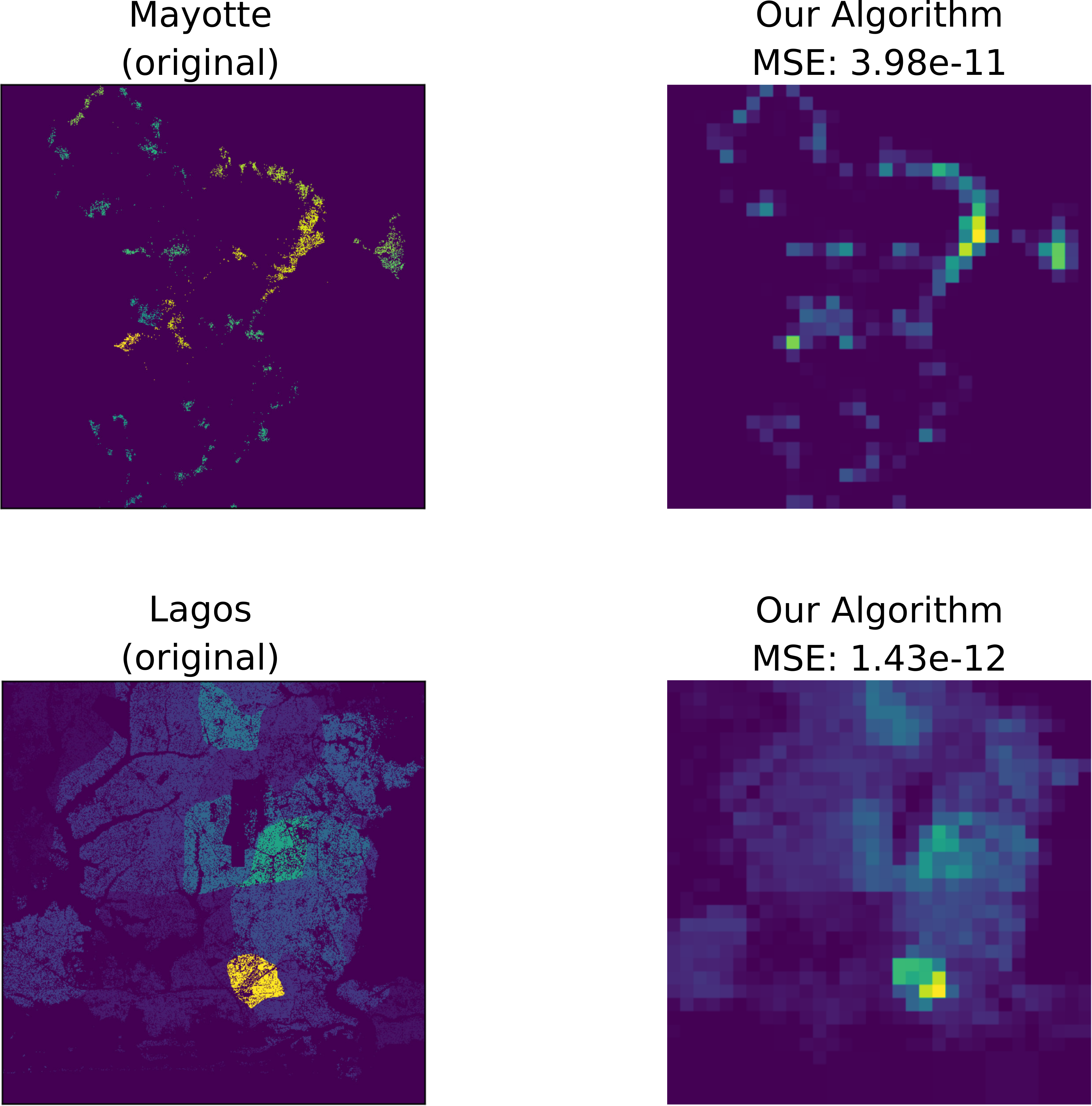}
    \caption{\textbf{Visual comparison of original (left) and differentially private heatmap (right) on two location datasets ($100k$ users sampled) using privacy budget $\varepsilon=1$.}}
    \label{fig:other}
\end{figure}

\paragraphbe{Our algorithm} Interactivity enables communication savings by only querying data at granularities with sufficient client density. Recall that our algorithm defines a privacy schedule proportional to the number of regions in every level so that the algorithm has clear stopping criteria preventing unnecessary communication with little utility and wasted budget. Furthermore, a threshold that adapts to the amount of applied noise allows reaching a fine level with a more accurate quaternary tree split, resulting in $7.88e^{{-}13}$ error which is only $1.7\%$ more than the original sampling error in a non-private algorithm. The algorithm achieves this with a communication cost (vector size) of $340$, which is orders of magnitude lower communication compared to the other private baselines. We further study hyperparameter tuning of our algorithm in Appendix~\ref{sec:ablation}.

\paragraphbe{Effect of other datasets} We pick maps from the Humanitarian Data Exchange for the archipelago of Mayotte, with population sparsely spread along the coasts, and for Lagos, Nigeria, with population spread across the whole map. For Manhattan, we pick areas of $1024 \times 1024$ cells. The selection of Lagos's map has a total of $9,578,296$ participants and Mayotte has $239,880$. 

Figure~\ref{fig:other} visually compares the density estimation accuracy of our algorithm. When using the same parameters as for the NYC map we reconstruct Mayotte's map with error $3.98e^{-11}$ and total communication of $929$ whereas an interactive fixed epsilon, non-adaptive threshold gives $4.45e^{-11}$ with communication $27,022$. For Lagos, our algorithm obtains error $1.43e^{-12}$ and $291$ overhead compared to $1.82e^{-12}$ with overhead $41,233$. 

\paragraphbe{Multiple locations per user} For this experiment we use a location check-in dataset from Tokyo~\cite{foursquare_dataset} since it encodes which locations were visited by the same user. In absence of sufficient temporal information, we simply treat user location data as a multiset with integer visits to each location. There are a total of $2,293$ users and $573,695$ locations. We preprocess the dataset into the grid of $64\times64$ and run our algorithm with different scaling $\gamma$ and different quantization, i.e. modulo clipping constraints. Recall that for the multi-location setting we generate noise with sensitivity $\gamma + d$ where $d$ represents dimensions of the reported vector. As baselines we run our algorithm without privacy constraints and a single-location version by simply picking the most visited location for each user. As we increase scaling we also increase the added noise and therefore can cause overflow. Table~\ref{tab:multi_loc} shows this effect, increasing the scaling factor first improves results until it overflows on clipping.

\begin{table}[]
\caption{Multiple locations per user extension. $\varepsilon=1$.}
\label{tab:multi_loc}
\centering
\begin{tabular}{lrrrrr}
Algorithm & $\gamma$ & clip      & MSE (e-7) & $L_1$ (e-2) & Comm \\
\midrule
Non-private    & $1$    & 16      & $1.2$  & $2.64$ &  $5,460$  \\
Single location& $1$    & 16      & $13.0$ & $9.7$  & $293$ \\ 
Multi-location & $1$    & 16      & $18.3$ & $11.6$ & $1,108$ \\
Multi-location & $10^2$ & 16      & $19.2$ & $10.8$ & $1,523$\\
Multi-location & $10^2$ & 32      & $16.9$ & $10.8$  & $1,331$ \\
Multi-location & \textbf{$10^4$} & \textbf{32}      & \textbf{7.7}  & \textbf{6.8}  & $1,746$\\
Multi-location & $10^6$ & 32      & $8.3$ & $7.9$ & $1,218$ \\
\bottomrule
\end{tabular}
\end{table}

\section{Discussion}

\paragraphbe{Communication and Computation costs for SecAgg}
Our analysis so far has not considered additional communication overhead imposed by the secure aggregation protocol. For $n$ clients, each inputting a vector of length $l$, the secure aggregation protocol from~\cite{bell2020secagg}
requires $O(\log n + l)$ communication and
$O(\log^2 n + l\log n)$ computation per client.
Note that the dominating term in both client communication
and computation is $l$. This is the case also in terms of concrete 
efficiency, as shown by Bell et al.~\cite{bell2020secagg}.
For this reason we simply use the value $l$ as 
a proxy for the costs incurred by secure aggregation, taking into account that this protocol easily scales, in terms of both server and clients computation/communication costs, to vectors of length $1M$
and over $100k$ clients.

\paragraphbe{Prior information and dynamic datasets} Algorithm~\ref{alg:dynamic} iteratively refines the prefix tree $\mathcal{T}$, spending some privacy budget along the way. Intuitively, if fewer iterations are required to reach the optimal $\mathcal{T}$, less privacy budget can be spent discovering it and more can be spent collecting data at this resolution. Since Algorithm~\ref{alg:dynamic} supports both leaf node splitting and collapsing, we can in theory begin with any prefix tree $\mathcal{T}$, rather than always beginning with only the quadtree root node. This gives us the flexibility to guess at the optimal $\mathcal{T}$ and reap accuracy gains or losses depending on whether the guess is closer or farther from the optimal $\mathcal{T}$ than the root node. We can thus apply any prior we may have about the spatial distribution under measurement.

The epidemiological topics described in Section~\ref{subsec:applications}, such as the spread of disease, the danger of traffic, and other public health patterns, are not, in general, static. In these and other applications, re-querying a distribution can provide value by capturing its temporal shifts. When re-querying a distribution, the previous result can act as an informative prior, and may produce real accuracy gains when used as the starting point $\mathcal{T}$ as described above. We leave for future work further exploration of how best to leverage prior information when exploring dynamic distributions.

\section{Conclusion}
\label{sec:conclusion}
We revisited the distributed differential privacy concept in light of recent advances in secure multiparty computation that allow computing secure vector sums with contributions from thousands of participants. We showed through an end-to-end design how this makes it feasible to generate differentially private heatmaps over location data at the metropolitan scale with contributions from millions of users. In particular, we designed an adaptive hierarchical histogram algorithm that  interactively refines the resolution of the heatmap to construct an efficient dense vector representation of the location space. It exploits interactivity by adapting resolution, stopping criteria, and privacy budget allocation based on the results of prior histogram queries. The results with public location datasets show that this adaptive approach generates heatmaps with high accuracy while significantly reducing the worst case client communication overhead of existing privacy-preserving baselines with comparable accuracy. While this paper has focused on location heatmaps, opportunities exist to generalize this approach to other sparse federated analytics settings where private data is sparsely distributed over a large domain.

Privacy is a multi-faceted challenge. While this paper has focused on a private heatmap aggregation mechanism, a complete solution should address privacy more holistically. For example, storing raw data on device and accessing only aggregates does not absolve the service provider of proper stewardship obligations. Individual data should still be treated securely with standard best practices such as encryption-at-rest, access control, finite lifetime, processed only with informed consent, and follow contextual integrity principles~\cite{nissenbaum2004privacy}. We hope, however, that this paper advances the toolkit for designing privacy-aware systems.

\section*{Acknowledgments}

At Cornell Tech, Bagdasaryan is supported in part by a Cornell Digital Life Initiative fellowship and an Apple Scholars in AI/ML fellowship.

\bibliographystyle{abbrv}
\bibliography{main}

\appendix

\section{Ablation Study}
\label{sec:ablation}

To understand how the different components of the algorithm contribute, Table~\ref{tab:ablationresults} presents other interactive algorithm variants. 

As a \textit{non-private version}, consider running the hierarchical histograms algorithm with no differential privacy noise (infinite privacy budget) and setting the threshold to $10$, achieving the same error as the non-interactive, non-private baseline with communication cost of $12,734$. While this is worse than the communication cost of sending location in the clear, it is far superior to the non-interactive one-hot encoding approaches, saving costs by not expanding nodes that had low counts on every level of the quadtree. 

A simple private but \textit{non-adaptive hierarchical histogram} algorithm simply splits the $\varepsilon=1$ privacy budget evenly among the quadtree levels of the non-private version. By using threshold $T=10$ we achieve similar error $8.81e^{{-}13}$ to non-interactive run but with only $43,450$ overhead. The increase in communication from the non-private interactive baseline is caused by additional splits due to high equal noise ($\varepsilon=0.1$) across levels.

Using our schedule and a fixed threshold results in $7.95e^{{-}13}$ error and expansion to only first 4 levels of the quadtree while also leaving most of the budget ($0.76$) to the last level, thus getting accurate final estimates. Furthermore, a threshold that adapts to the amount of applied noise allows reaching a finer level with a more accurate quadtree split, resulting in an error of $7.88e^{{-}13}$.

\subsection{Additional Measures of Error}
\label{subsec:more_metrics}
As described in Section~\ref{subsec:applications}, different measures of error are appropriate for different applications. Table~\ref{tab:ablationresults} demonstrates that our algorithm is not overtuned to MSE in particular; it includes several other measures of error which have historically been recommended for evaluating differences between epidemic distributions~\cite{tabataba2017framework}: $L_1$ distance (or absolute error), mean absolute percentage error (MAPE, using the minimum location value as the denominator when the true value is zero), symmetric absolute percentage error (sMAPE), and mean arctangent absolute percentage error (MAAPE). For absolute measures of error, our algorithm's increase in error compared to the non-private baseline ranges from $0\%$ for MAAPE to $1.9\%$ for $L_1$ distance, while for relative measures of error (which are far more sensitive to noise in low-population regions), it is just $8.7\%$ for sMAPE $11\%$ for MAPE. That our algorithm performs so close to the non-private baseline on all of these error measures suggests that it is broadly applicable even when the application does not call precisely for MSE.

\begin{table*}[]
\caption{Ablation study of our algorithm with $10k$ users (no dropout) and privacy budget $\varepsilon=1$.}
\label{tab:ablationresults}
\centering
\begin{tabular}{lrrrrrr}
Algorithm           & MSE (e-13) & $L_1$ (e-3) & MAPE & sMAPE & MAAPE & Comm \\
\midrule
No noise (T=10, $\varepsilon=\infty$) & $7.75$ & $1.02$ & $3.10$ & $0.92$ & $0.79$ & $12,734$ \\
Fixed $\varepsilon$, non-adaptive ($T=10$)      & $8.81$ & $1.10$ & $3.94$ & $1.12$ & $0.80$ & $43,450$  \\
Scheduled $\varepsilon$, non-adaptive ($T=10$) & $7.95$ & $1.05$ & $3.46$ & $1.09$ & $0.79$  & $340$ \\
Scheduled $\varepsilon$, adaptive $T$ & $7.88$ & $1.05$ & $3.43$ & $1.00$ & $0.79$  & $340$ \\

\bottomrule
\end{tabular}
\end{table*}

\begin{figure}[t]
    \centering
    \includegraphics[width=1.0\linewidth]{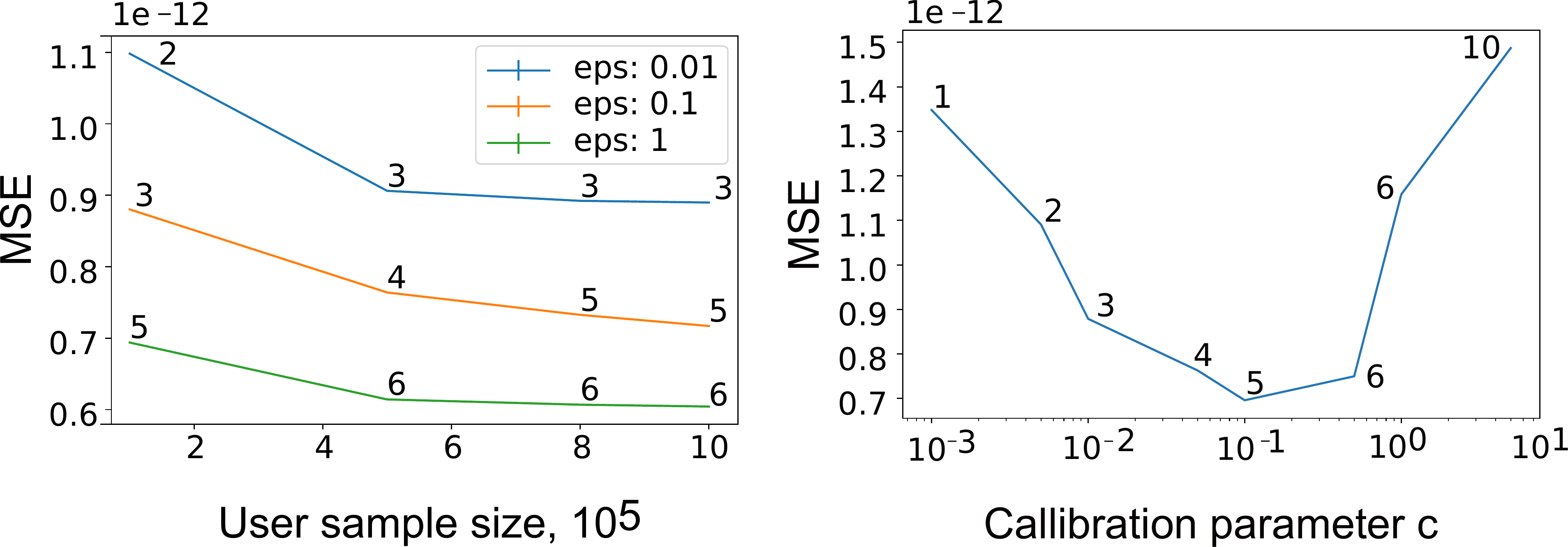}
    \caption{ Sensitivity analysis. Algorithm runs more levels as sample size (left) or calibration parameter $c$ (right) increases, but performance only improves with larger sample size.}
    \label{fig:hyperparams}
\end{figure}

\begin{figure*}[ht!]
    \centering
    \includegraphics[width=0.95\linewidth]{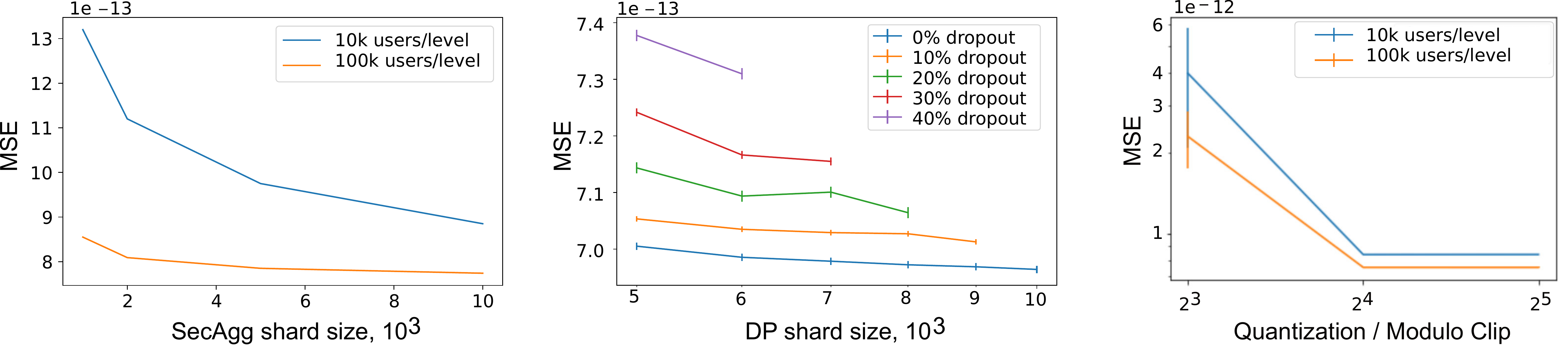}
    \caption{\textbf{Secure aggregation parameters have profound impact on performance.} Large shard size (left) and modulo clip (right) improve performance, while adjusting the DP shard size to dropout achieves DP guarantees with slight performance drop.}
    \label{fig:ablation}
\end{figure*}

\subsection{Sensitivity Analysis}

\paragraphbe{Effect of a sample size} Sampling more users into the algorithm increases the signal but also amplifies the differentially private noise. 
 However, accumulated noise would have standard deviation $\sigma_{aggr} = \sqrt{k*\sigma^2}$ for $k$ parallel SecAgg shards with geometric noise $\sigma$. For example, for a SecAgg shard of size $10,000$, sampling $100,000$ users will result in only $\sqrt{10}$ more noise than sampling $10,000$ users while the signal can grow linearly. Figure~\ref{fig:hyperparams} (left) shows that the algorithm benefits from large sample sizes even under very strict privacy budgets. 
 
 \paragraphbe{Calibration parameter} As discussed in Section~\ref{sec:algorithm} calibration hyperparameter $c$ from  in Equation~\ref{eq:sigma} determines target standard deviation of the differential privacy noise. In other words it represents our estimates of the data distribution such as how concentrated is the population on the map. Figure~\ref{fig:hyperparams} (right) shows that high values of this parameter result in more levels explored by the algorithm and we achieve the highest performance when $c=0.1$. Finding appropriate calibration parameter is a non-DP operation and can be done using public data for that location or other, similar cities. 



\subsection{Effect of Secure Aggregation Parameters}
\label{subsec:secagg_effect}
To support real-world deployment our approach needs to be scalable and robust to user dropout. Any secure aggregation framework has its own scalability limitations, we evaluate our approach under these constraints and report performance.

\paragraphbe{SecAgg shard size} Figure~\ref{fig:ablation} (left) illustrates that larger SecAgg shard size leads to lower errors as it reduces excess differential privacy noise. The graph shows the average 
mean square error for density estimation on the New York dataset over 
a fixed quadtree depth $3$ with $\varepsilon=0.1$ per each level.
Still, using a sample size larger than the number of users that can be queried in one secure aggregation shard can improve density estimation accuracy. For example, at a shard size of 10,000, a sample size of 100,000 leads to lower error than the sample size of 10,000. 


\paragraphbe{Analyzing dropouts} In Fig.~\ref{fig:ablation} (center), we see that increasing the DP shard size (up to the secure aggregation shard size $S_{max}$) positively impacts the accuracy of the algorithm. Note that DP shard size corresponds to $\alpha$ of the  P\'{o}lya random variables (see Section~\ref{sec:systemdesign}). However, if the number of participants remaining in the secure aggregation shard is lower than the DP shard size  then the result of the algorithm cannot achieve the target DP budget. Fig.~\ref{fig:ablation} shows that with up to 20\% dropout rate and DP size of $8000$, we get only 15\% error increase over no user dropout.

The graph shows that even when overcompensating for significant fraction of participants dropping out during the sum computation, for example due to loss of network connectivity, the resulting mean square error remains far below that of existing techniques such as local differential privacy techniques.

\paragraphbe{Quantization of vectors} Secure Aggregation operates on a finite set of integers but the proposed discrete Laplace mechanism achieves differential privacy by adding integer noise, which, in theory, could be unbounded. Thus, it is important to understand how to set the range for secure aggregation to avoid overflowing with high probability as this can significantly impact the performance. Figure~\ref{fig:ablation} (right) shows that after 16-bit integers the error remains the same, i.e. there is no overflow in the results and all the error is due to differential privacy noise.

\section{Adaptive grid and confidence intervals}
\label{sec:adaptive}


Some applications might utilize more information than users' locations to build a map. For example, an epidemiologist may wish to measure the prevalence of a disease within a population; in this case users can send infection test results along with their locations. For each leaf $l_i$ in the tree we can add data structure $data_i$ that stores the result status.

The user then reports not just the leaf but also the exact point in the data structure represented. We can represent the data structure similarly as a one-hot vector, retaining sensitivity $1$ for the reported vector, or we can adjust the sensitivity (and therefore differential privacy noise) according to some other maximum contribution.

\begin{figure}[t!]
    \centering
    \includegraphics[width=1.0\linewidth]{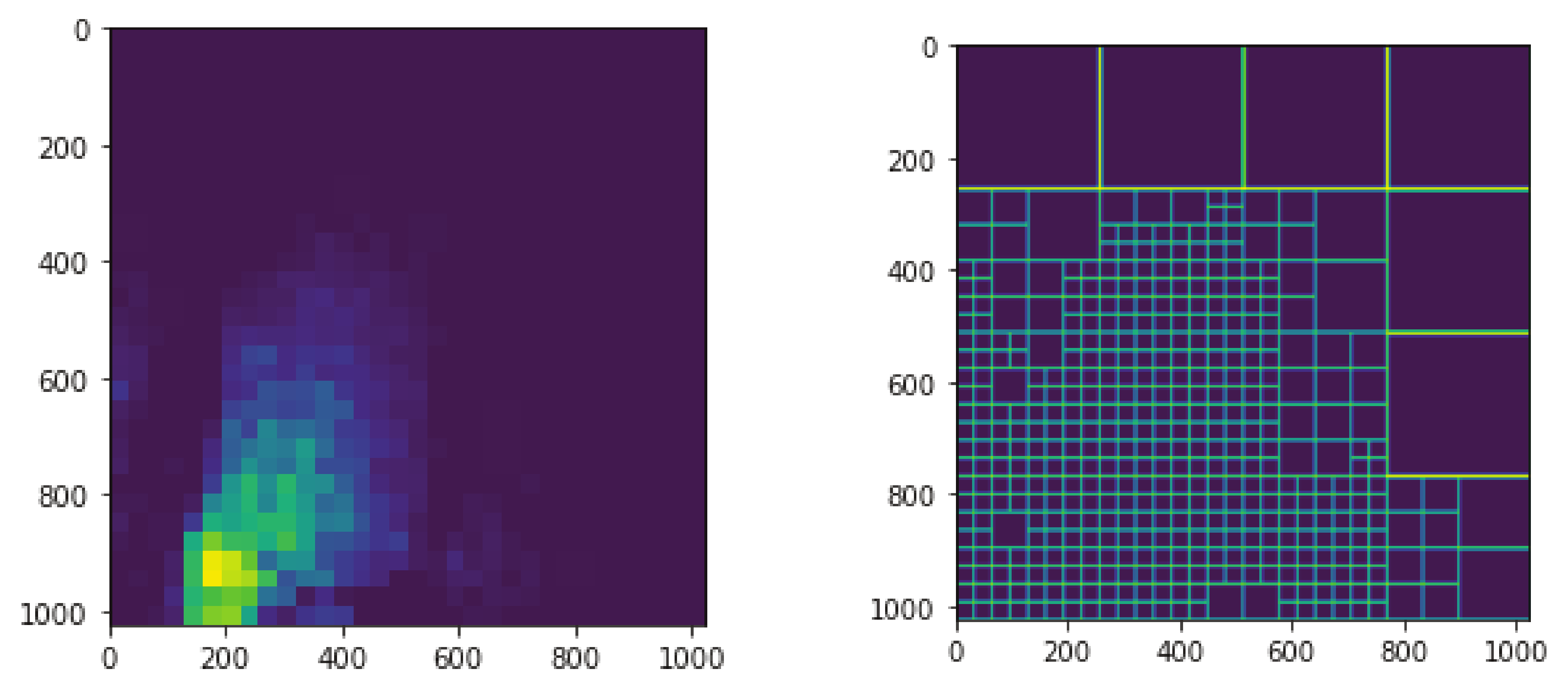}    \caption{\textbf{Adaptive grid}.}
    \label{fig:additional_data}
\end{figure}

Note that a data structure of size $|data|$ for a tree with $T$ leaves will occupy $|data| \cdot T$ space.  It may be possible to run the hierarchical algorithm over the auxiliary data to reduce size of $data$ but we only consider here the small-dimensional case (such as the single-dimensional case of sending an `infected' bit). Algorithm~\ref{alg:1} can support auxiliary data by modifying the ClientUpdate method to encode the data into the vector.

Depending on the precise application, the UpdateTree function from Algorithm~\ref{alg:dynamic} can be altered to split or collapse node $l_i$ based on the auxiliary data aggregated in structure instead or in addition to that node's user count. In the disease prevalence example, we can split or collapse based on the size of the confidence interval of the proportion estimate. 

\paragraphbe{Picking the metric} Along with MSE we consider another metric that is important when the data will be used to guide high-cost or high-consequence action: confidence intervals on proportion of positive cases in every region. This metric is critical for public health and other sensitive applications where the cost and/or consequence of action imposes some minimum confidence level to consider the data actionable, such as when the map will be used to guide distribution of scarce resources like medical supplies or money for infrastructure.

\paragraphbe{Additional data} Our algorithm can support collection of auxiliary location-associated data with minimal tweaks. Fig.~\ref{fig:additional_data} shows our algorithm's output for the proportion estimation extension after setting an auxiliary 'infected' bit according to a Gaussian 2D kernel with the center at $(200, 900)$ that approximately corresponds to Lower Manhattan and assigning the value $1$ to points within the kernel at $0$.



\end{document}